\documentclass[journal,onecolumn]{IEEEtran}
\IEEEoverridecommandlockouts

\usepackage{cite,setspace}
\usepackage{graphicx}
\usepackage{psfrag}
\usepackage{subfigure}
\usepackage{url}
\usepackage{stfloats}
\usepackage{amsfonts,amssymb,amsmath,bm,paralist,theorem,cite,ifthen,color}
\usepackage{array}
\usepackage{caption}
\usepackage{calc}
\usepackage{subfig}
\usepackage{longtable}
\usepackage{stfloats}  
\usepackage{algpseudocode}
\usepackage{algorithm}
\usepackage{algorithmicx}
\usepackage{epstopdf}
\usepackage{pdfsync}
\usepackage{graphics,booktabs,color,epsfig,subfigure}

\usepackage{float}
\restylefloat{table,figure}

\def\cV{\mathcal{V}}
\def\cD{\mathcal{D}}
\def\cL{\mathcal{L}}
\def\cF{\mathcal{F}}
\def\cS{\mathcal{S}}
\def\cV{\mathcal{V}}
\def\cK{\mathcal{K}}
\def\bz{\mathbf{z}}
\def\br{\mathbf{r}}
\def\bx{\mathbf{x}}
\def\cB{\mathcal{B}}

\def\cN{\mathcal{N}}

\newtheorem{thm}{Theorem~}

\begin{document}
%
\title{Network Slicing for Service-Oriented Networks Under Resource Constraints}
%
%
%

\author{Nan~Zhang, Ya-Feng~Liu, Hamid~Farmanbar, Tsung-Hui~Chang, Mingyi~Hong, and Zhi-Quan~Luo
\thanks{This work is supported by NSF, grant number CCF-1526434, and by NSFC, grant number 61571384.}
\thanks{N.~Zhang is with the School of Mathematical Sciences, Peking University, China. Email: zhangnan625@pku.edu.cn}
\thanks{Y.-F.~Liu is with the State Key Laboratory of Scientific and Engineering Computing, Institute of Computational
Mathematics and Scientific/Engineering Computing, Academy of Mathematics and Systems Science, Chinese Academy of Sciences, Beijing, China. Email: yafliu@lsec.cc.ac.cn}
\thanks{H.~Farmanbar is with Huawei Canada Research Center, Ottawa, Canada. Email: hamid.farmanbar@huawei.com}
\thanks{M.~Hong is with the Department of Electrical and Computer Engineering, University of Minnesota, USA. Email: mhong@umn.edu}
\thanks{T.-H.~Chang and Z.-Q.~Luo are with Shenzhen Research Institute of Big Data, and the Chinese University of Hong Kong, Shenzhen, China. Emails: changtsunghui@cuhk.edu.cn,~luozq@cuhk.edu.cn}

}

%
%

{}
%



\maketitle

\begin{abstract}
To support multiple on-demand services over fixed communication networks, network operators must allow flexible customization and fast provision of their network resources. One effective approach to this end is network virtualization, whereby each service is mapped to a virtual subnetwork providing dedicated on-demand support to network users. In practice, each service consists of a prespecified sequence of functions, called a service function chain (SFC), while each service function in a SFC can only be provided by some given network nodes. Thus, to support a given service, we must select network function nodes according to the SFC and determine the routing strategy through the function nodes in a specified order.
A crucial network slicing problem that needs to be addressed is how to optimally localize the service functions in a physical network as  specified by the SFCs, subject to link and node capacity constraints.
In this paper, we formulate the network slicing problem as a mixed binary linear program and establish its strong NP-hardness. Furthermore, we propose efficient penalty successive upper bound minimization (PSUM) and PSUM-R(ounding) algorithms, and two heuristic algorithms to solve the problem. Simulation results are shown to demonstrate the effectiveness of the proposed algorithms.
\end{abstract}


%
\IEEEpeerreviewmaketitle

\section{Introduction}\label{Sec:intro}

\IEEEPARstart{T}{oday's} communication networks are expected to support multiple services with diverse characteristics and requirements. To provide such services efficiently, it is highly desirable to make the networks agile and software reconfigurable. Network function virtualization (NFV) \cite{chiosinetwork,Mijumbi2015} is an important technology to achieve this goal, which virtualizes network service functions so that they are not restricted to the dedicated physical devices.
Different from the traditional networking where service functions are assigned to special network hardware, NFV enables service operators to flexibly deploy network functions and service providers to intelligently integrate a variety of network resources owned by different operators to establish a service customized virtual network (VN) for each service request. In practice, each service consists of a predefined sequence of service functions, called a service function chain (SFC) \cite{SFC2015}. Meanwhile, a service function can only be provided by certain specific nodes, called NFV-enabled nodes. As all of the VNs share a common resource pool, we are led to the problem of network resource allocation to meet diverse service requirements, subject to the capacity constraints at NFV-enabled nodes and at network links.

Recently, reference \cite{Zhang2015} proposed a novel 5G wireless network architecture MyNET and an enabling technique called SONAC (Service-Oriented Virtual Network Auto-Creation). In SONAC, there are two key components: software defined topology (SDT) and software defined resource allocation (SDRA). SDT determines a VN graph and a VN logical topology for each service request. The determination of the VN logical topology, also called VN embedding, locates the virtual service functions onto the physical NFV-enabled infrastructures so that each function in the corresponding SFC is instantiated.
SDRA maps the logical topology to physical network resources, including both communication and computational resources. 

In software defined networks, centralized control enables joint VN embedding and SDRA, a problem called network slicing. Specifically, network slicing controls the flow routing such that each flow gets processed at NFV-enabled nodes in the order of service functions defined in the corresponding SFC. There are some works related to the network slicing problem \cite{Jiang2012,Narayana2013,Xu_2013,Li2015,Gushchin2015,Li2016,Charikar2016,Abbasi2015,Chua2016,Zhang2015a,Ghaznavi2016,Addis2015,ServiceMapping}.
Reference \cite{Zhang2015a} considered a simplified problem where there is a single function in each SFC and a single path for each service, and solved the formulated problem approximately.
References \cite{Jiang2012,Narayana2013} simplified ``routing" by either considering only one-hop routing or selecting paths from a predetermined path set. Reference \cite{Gushchin2015} considered the so-called consolidated middleboxes where a flow could receive all the required functions. It proposed a two-stage heuristic algorithm to route each flow through a single associated NFV-enabled node. Such formulation is not applicable to the case where each SFC contains multiple functions which need to be instantiated in sequence at NFV-enabled nodes. An important common assumption in \cite{Narayana2013,Xu_2013,Li2016,Charikar2016,Abbasi2015,Chua2016} is that the instantiation of a service function for a traffic flow can be split over multiple NFV-enabled nodes. The service splitting assumption significantly simplifies the optimization problem since no binary variable is needed in the problem formulation. However, service splitting would result in high coordination overhead in practice, especially when the number of service requests is large. Reference \cite{Ghaznavi2016} allowed service splitting in a different way by assuming that a service function can be instantiated in the form of multiple instances of virtual network functions (with different throughputs and resource demands) at multiple nodes.
References \cite{Li2015,Addis2015,ServiceMapping} did not allow service splitting and assumed the data of any flow are processed in only one NFV-enabled node for any function in the chain. They solved their problems by either existing integer optimization solvers or heuristic algorithms that may lead to violations of resource constraints.

In this paper, we consider the network slicing problem with practical constraints, where a set of service requests are simultaneously processed and routed. Our considered problem differs from most of the aforementioned works in that we allow traffic flows to be transmitted on multiple paths and require that the data of any flow are processed by only one NFV-enabled node for any function in the corresponding function chain.
We formulate the problem as a mixed binary linear program and show that checking the feasibility of this problem is strongly NP-hard in general. Moreover, we propose an efficient penalty successive upper bound minimization (PSUM) algorithm, a PSUM-R algorithm, and two heuristic algorithms for this problem. The simulation results show that the proposed PSUM and PSUM-R algorithms can find a near-optimal solution of the problem efficiently.

\section{System Model and Problem Formulation}\label{Sec:model}
In this section, we introduce the model and present a formulation of the network slicing problem.

Consider a communication network represented by a graph $\mathcal{G}=(\mathcal{V},\mathcal{L})$, where $\mathcal{V}=\{i\}$ is the set of nodes and $\cL=\{(i,j)\}$ is the set of directed links. Denote the subset of function nodes (NFV-enabled nodes) that can provide a  service function $f$ as $V_f$. Each function node $i$ has a known computational capacity $\mu_i$, and we assume that processing one unit of data flow requires one unit of computational capacity.
Suppose that there are $K$ data flows, each requesting a distinct service in the network. The requirement of each service is given by a service function chain $\cF(k)$, consisting of a set of functions that have to be performed in the predefined order by the network.
Different from the aforementioned works \cite{Li2016,Charikar2016,Abbasi2015,Chua2016}, we require that each flow $k$ receives each service function in $\cF(k)$ at exactly one function node, i.e., all data packets of flow $k$ should be directed to the same function node to get processed by a service function,
so that there does not exist coordination overhead caused by service splitting in practice.
Notice that this requirement does not prevent a common function in different SFCs from being served by different nodes. 
The source-destination pair of flow $k$ is given as ($S(k),D(k)$), and the arrival data rate of flow $k$ is given as $\lambda(k)$.
The network slicing problem is to determine the routes and the rates of all flows on the routes while satisfying the SFC requirements and the capacity constraints of all links and function nodes. 

Let $r_{ij}(k)$ be the rate of flow $k$ over link $(i,j)$. The capacity of link $(i,j)$ is assumed to be $C_{ij}$ which is a known constant. This assumption is reasonable when the channel condition is stable during the considered period of time.
The total flow rates over link $(i,j)$ is then upper bounded by $C_{ij}$, i.e., 
\begin{equation}\label{link_capa}
  \sum_{k=1}^Kr_{ij}(k)\leq C_{ij},~\forall\,(i,j)\in \cL. 
\end{equation}

To describe VN embedding, we introduce binary variables $x_{i,f}(k)$ which indicate whether or not node $i$ provides function $f$ for flow $k$ (i.e., $x_{i,f}(k)=1$ if node $i$ provides function $f$ for flow $k$; otherwise $x_{i,f}(k)=0$).
To ensure that each flow $k$ is served by exactly one node for each $f\in \cF(k)$, we have the following constraint 
\begin{equation}\label{1norm}
  \sum\limits_{i\in V_f} x_{i,f}(k)=1,~\forall\,f\in \cF(k),~\forall\,k. 
\end{equation}

\begin{table}[t]
  \centering
  \caption{Summary of Notations}\vspace{-0.5em}
  \begin{tabular}{c|l}
     \hline
     $V_f$ & subset of nodes that can provide function $f$ \\
     \hline
      $\cF(k)$ & service function chain of flow $k$\\
     \hline
     $\lambda(k)$ & arrival data rate of flow $k$\\
     \hline
     $(S(k),D(k))$ & source-destination pair of flow $k$ \\
     \hline
     $C_{ij}$ & communication capacity of link $(i,j)$\\
     \hline
     $\mu_i$ &  computational capacity of node $i$\\
     \hline
     $x_i$    & binary variable indicating whether or not \\
     & node $i$ is active\\
     \hline
     $x_{i,f}$ & binary variable indicating whether or not \\
     & node $i$ provides function $f$ \\
     \hline
     $x_{i,f}(k)$ & binary variable indicating whether or not  \\
     & node $i$ provides function $f$ for flow $k$\\
     \hline
     $r_{ij}(k)$ & rate of flow $k$ over link $(i,j)$\\
     \hline
     $r_{ij}(k,f)$ & rate of virtual flow $(k,f)$ over link $(i,j)$\\
     \hline
   \end{tabular}\vspace{-0.5em}
  \label{table:notation}
\end{table}

We assume that each function node provides at most one function for each flow: 
\begin{equation}\label{atmost1}
  \sum\limits_{f\in \cF(k)} x_{i,f}(k)\leq 1, ~\forall\,k,~\forall\,i. 
\end{equation}
This assumption is without loss of generality.
This is because, if a function node can provide multiple services for a flow, we can introduce virtual nodes such that each virtual node provides one function for the flow and all these virtual nodes are connected with each other.

\begin{figure}[h]
  \centering
  \includegraphics[width=7.8cm]{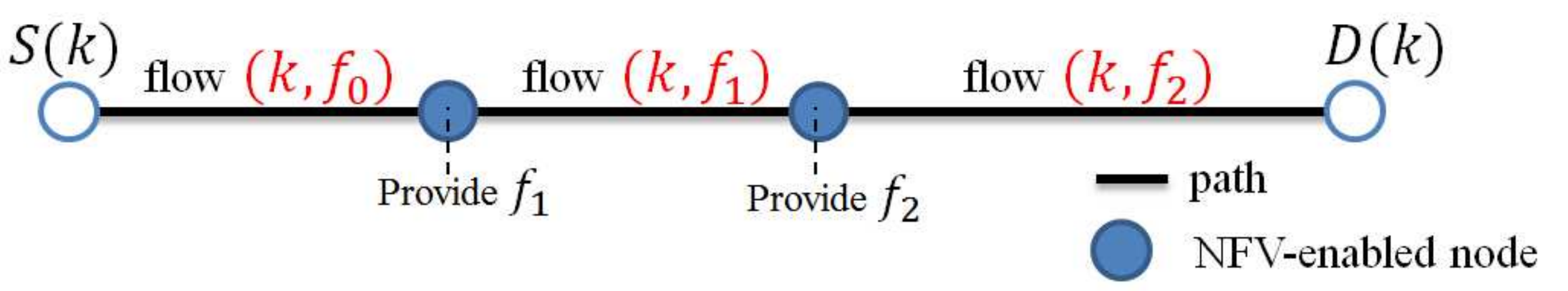}\vspace{-0.3em}
  \caption{\small An illustration of the virtual flow.}\label{virtualflow}\vspace{-0.8em}
\end{figure}

Suppose that the function chain of flow $k$ is $\cF(k)=(f^k_1\rightarrow f^k_2\rightarrow\dots\rightarrow f^k_n)$.
To ensure flow $k$ goes into the function nodes in the prespecified order of the functions in $\cF(k)$,
we introduce new virtual flows labelled $(k,f)$: flow $k$ just after receiving the service function $f$, and denote as $(k,f^k_0)$ the flow $k$ just coming out of the source node $S(k)$ without receiving any service function.
See Fig. \ref{virtualflow} for an illustration.
Let $r_{ij}(k,f)$ be the rate of flow $(k,f)$ over link $(i,j)$.
Since each function node provides at most one function for each flow (as shown in \eqref{atmost1}), the following flow conservation constraints must hold for all nodes $i$ and $s=1,\dots,n$: 
\begin{equation}\label{flow_cons1}
\hspace{-0.4em}\lambda(k)x_{i,f_s}(k)\!=\!\sum\limits_{j:(j,i)\in \cL}r_{ji}(k,f_{s-1})\!-\!\sum\limits_{j:(i,j)\in \cL}r_{ij}(k,f_{s-1}), 
\end{equation}
\begin{equation}\label{flow_cons2}
\lambda(k)x_{i,f_s}(k)\!=\!\sum\limits_{j:(i,j)\in \cL}r_{ij}(k,f_s)\!-\!\sum\limits_{j:(j,i)\in \cL}r_{ji}(k,f_s), 
\end{equation}
\begin{equation}\label{flow_cons3}
\sum\limits_{j:(S(k),j)\in \cL}r_{S(k)j}(k,f_0)=\lambda(k),
\end{equation}
\begin{equation}\label{flow_cons4}
  \sum\limits_{j:(j,D(k))\in \cL}r_{jD(k)}(k,f_n)=\lambda(k), 
\end{equation}
where \eqref{flow_cons1} and \eqref{flow_cons2} imply that if $x_{i,f_s}=1$, then flow ($k,f_{s-1}$) going into node $i$ and flow ($k,f_s$) coming out of node $i$ both have rate $\lambda(k)$; otherwise each virtual flow $(k,f_s)$ coming out of node $i$ and going into node $i$ should have the same rate; \eqref{flow_cons3} and \eqref{flow_cons4} ensure that flow $k$ coming out of $S(k)$ and going into $D(k)$ both have rate $\lambda(k)$.
These constraints guarantee that each flow $k$ gets served at function nodes in the order prespecified by $\cF(k)$ and with the required data rate $\lambda(k)$.

By the definitions of $r_{ij}(k,f)$ and $r_{ij}(k)$, we have 
\begin{equation}\label{virtualsum}
  r_{ij}(k)=\sum_{f\in \cF(k)\cup{\{f^k_0\}}} r_{ij}(k,f),~\forall\,k,~\forall\,(i,j)\in\cL. 
\end{equation}

Since processing one unit of data flow consumes one unit of computational capacity, the node capacity constraint can be expressed as 
\begin{equation}\label{node_capa}
\sum_f\sum_k\lambda(k) x_{i,f}(k)\leq \mu_{i}, ~\forall\,i. 
\end{equation}

Now we present our problem formulation of network slicing to minimize the total link flow in network: 
\begin{equation}\label{prob}
\begin{array}{ll}
\hspace{-0.8em} \min\limits_{\br,\bx}
&  g(\br)=\sum\limits_{k,(i,j)}r_{ij}(k)\\
\hspace{-0.8em} \mbox{s.t.}&\eqref{link_capa}-\eqref{node_capa},\\
\hspace{-0.8em} & r_{ij}(k)\geq 0, ~\forall\,k,~\forall\,(i,j)\in\cL,\\
\hspace{-0.8em} & r_{ij}(k,f)\geq 0,~\forall\,f\in\cF(k),\forall\,k,\forall\,(i,j)\in \cL, \\
\hspace{-0.8em} & x_{i,f}(k)\in \{0,1\},~\forall\,i\in V_f,\forall\,f\in \cF(k),\forall\,k, 
\end{array}
\end{equation}
where $\br:=\bigl( \{r_{ij}\},\{r_{ij}(k,f)\}\bigr),\, \bx:=\{x_{i,f}(k)\}$.
The objective function $g(\br)=\sum_{k,(i,j)}r_{ij}(k)$ is set to avoid cycles in choosing routing paths.
There can be other choices of objective functions, such as the cost of the consumed computational resources and the number of activated function nodes.

Problem \eqref{prob} is a mixed binary linear program which turns out to be strongly NP-hard.
The proof is based on a polynomial time reduction from the 3-dimensional matching problem \cite{Karp1972}.
\vspace{-0.7em}

\begin{thm}\label{thm1}
   Checking the feasibility of problem \eqref{prob} is strongly NP-complete, and thus solving problem \eqref{prob} itself is strongly NP-hard. \vspace{-0.2em}
\end{thm}
We give the proof of Theorem \ref{thm1} in Appendix A.

\section{Proposed PSUM and PSUM-R Algorithms}\label{sec:alg}

Since problem \eqref{prob} is strongly NP-hard, it is computationally expensive to solve it to global optimality.
In this section, we propose efficient PSUM and PSUM-R algorithms to solve it approximately.
The basic idea of our proposed PSUM algorithm is to relax the binary variables in problem \eqref{prob} and add penalty terms to the objective function to induce binary solutions. The PSUM-R algorithm combines PSUM and a rounding technique so that a satisfactory solution can be obtained more efficiently than PSUM.

\subsection{PSUM Algorithm}\label{subsec:psum}

Notice that problem \eqref{prob} becomes a linear program (LP) when we relax the binary variables to be continuous.
Problem \eqref{prob} and its LP relaxation are generally not equivalent (in the sense that the optimal solution of the LP relaxation problem may not be binary).
The following Theorem \ref{thm2} provides some conditions under which the two problems are equivalent. 
The proof of Theorem \ref{thm2} is given in Appendix B.
\begin{thm}\label{thm2}
  Suppose $\mu_i\geq \bar\mu$ for all $i$, and $C_{ij}\geq \bar C$ for all $(i,j)$, where
   \begin{equation}\label{bounds}
   \bar\mu=\sum_{k=1}^K\lambda(k),~\bar C=\sum_{k=1}^K\lambda(k)(|\cF(k)|+1),
   \end{equation}
     and $|\cF(k)|$ denotes the number of functions in $\cF(k)$.
     Then the LP relaxation of problem \eqref{prob} always has a binary solution of \{$x_{i,f}(k)$\}.
     Moreover, the lower bounds in \eqref{bounds} are tight in the sense that there exists an instance
     of problem \eqref{prob} such that its LP relaxation problem does not have a binary solution of \{$x_{i,f}(k)$\} if one of the lower bounds is violated. \vspace{-0.3em}
\end{thm}

Theorem \ref{thm2} suggests that, if the link and node capacity are sufficiently large, then problem \eqref{prob} and its LP relaxation problem (which relaxes the binary variables to be continuous) are equivalent.
Moreover, if the link and node capacity are fixed, problem \eqref{prob} becomes more difficult to solve as the number of flows and the number of functions in the SFC increase, because the lower bounds in Theorem \ref{thm2} will be more likely violated.

To solve the general problem \eqref{prob}, our basic idea is to add an $L_p$ penalty term to the objective of the LP relaxation problem of \eqref{prob} to enforce the relaxed variables end up being binary.

\begin{figure}[h]
  \centering
  \includegraphics[width=5cm]{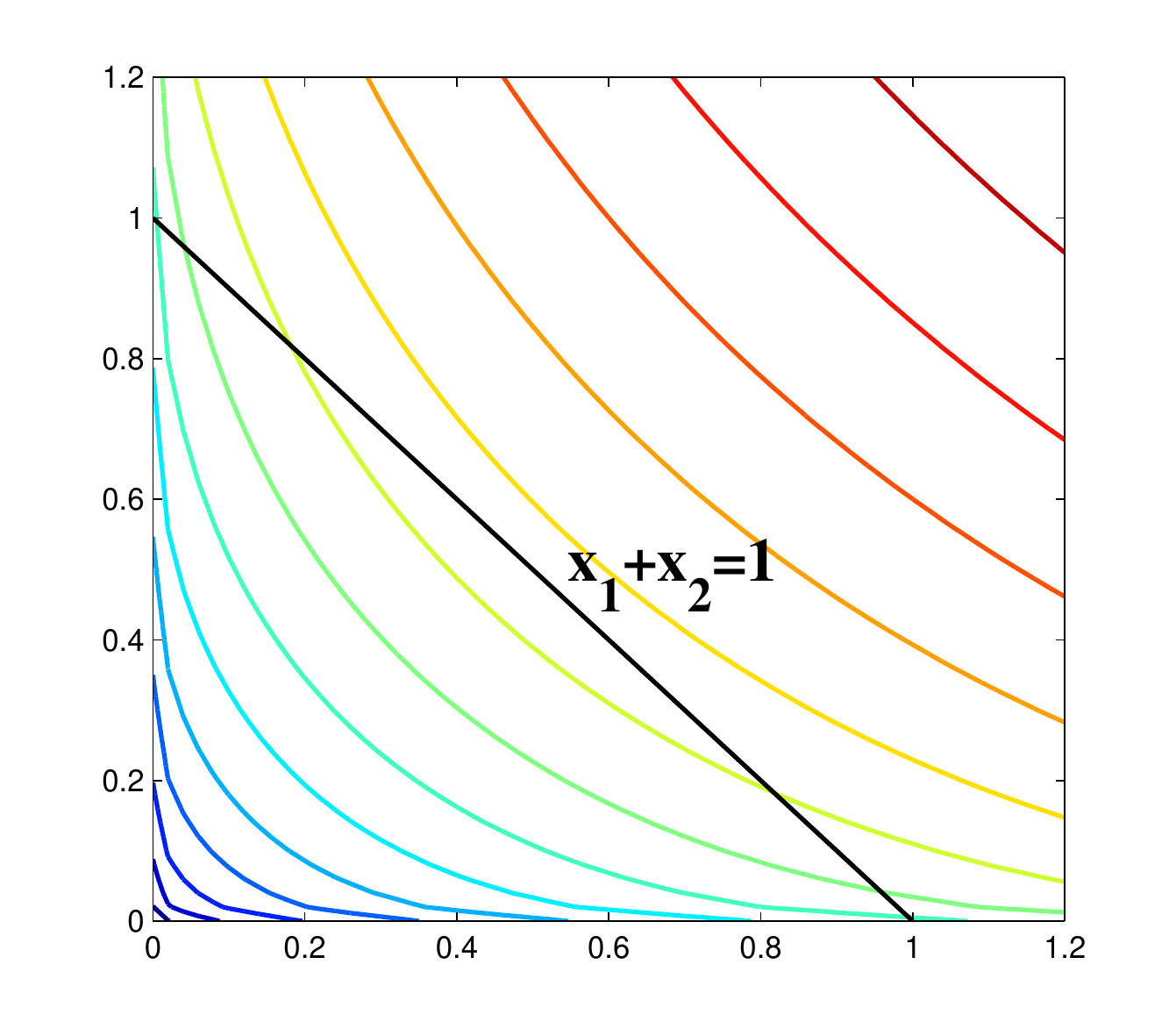}\vspace{-0.8em}
  \caption{\small Example: $\min~ \sqrt{x_1}+\sqrt{x_2}~ \textrm{subject to}~ x_1+x_2=1,x_i\geq0,i=1,2$, the optimal solution is $\mathbf{x}^*=(0,1),\,(1,0)$. } \vspace{-1em}
  \label{pnorm_example}
\end{figure}

Let $\bx_{f}(k)=\{x_{i,f}(k)\}_{i\in V_f}$. Then, we can rewrite
\eqref{1norm} as 
\begin{equation}\label{1norm_2}
  \|\bx_f(k)\|_1=1,~\forall\,f\in \cF(k),~\forall\,k.
\end{equation}
We have the following fact \cite{Liu2015}.\\
\textbf{Fact:} For any $k$ and any $f\in \cF(k)$, consider 
\begin{equation}\label{penalty_moti}
\begin{array}{ll}
\min\limits_{\bx_f(k)} & \|\bx_{f}(k)+\epsilon \mathbf{1}\|_p^p:=\sum\limits_{i\in V_f} (x_{i,f}(k)+\epsilon)^p \\
\mbox{s.t.}& \|\bx_{f}(k)\|_1=1,\\
& x_{i,f}(k)\in [0,1],~\forall\,i\in V_f, 
\end{array}
\end{equation}
where $p\in(0,1)$ and $\epsilon$ is any nonnegative constant. The optimal solution of problem \eqref{penalty_moti} is binary, that is, only one element is one and all the others are zero (see an example in Fig. \ref{pnorm_example}), and its optimal value is
$c_{\epsilon,f}:=(1+\epsilon)^p+(|V_f|-1)\epsilon^p$. Moreover, the objective function in problem \eqref{penalty_moti} is differentiable with respect to each element $x_{i,f}(k)\in[0,1]$ when $\epsilon>0$. 

Motivated by the above fact, we propose to solve the following penalized problem 
\begin{equation}\label{prob_penalty}
\begin{array}{ll}
\hspace{-0.8em}\min\limits_{\br,\bx}
& g(\br)+\sigma P_{\epsilon}(\bx)\\
\hspace{-0.8em} \mbox{s.t.}&\eqref{link_capa}-\eqref{node_capa},\\
\hspace{-0.8em} & r_{ij}(k)\geq 0, ~\forall\,k,~\forall\,(i,j)\in\cL,\\
\hspace{-0.8em} & r_{ij}(k,f)\geq 0,~\forall\,f\in\cF(k),\forall\,k,\forall\,(i,j)\in \cL, \\
\hspace{-0.8em} & x_{i,f}(k)\in [0,1],~\forall\,i\in V_f,\forall\,f\in \cF(k),\forall\,k, \vspace{-0.2em}
\end{array}
\end{equation}
where $\sigma>0$ is the penalty parameter, and \vspace{-0.1em}
\begin{equation}\label{pnorm_penalty}
 P_{\epsilon}(\bx)= \sum\limits_{k}\sum\limits_{f\in \cF(k)} \bigl( \|\bx_{f}(k)+\epsilon\mathbf{1}\|_p^p-c_{\epsilon,f}\bigr). 
\end{equation}

For ease of presentation, we define $\bz=(\br,\bx)$ and $g_\sigma(\bz)=g(\br)+\sigma P_\epsilon(\bx)$.
The following Theorem \ref{thm3} reveals the relationship between the optimal solutions of problems \eqref{prob} and \eqref{prob_penalty}. The proof is relegated to Appendix \ref{sec:penalProof}. 

\begin{thm}\label{thm3}
 For any fixed $\epsilon>0$, let $\bz^k$ be a global minimizer of problem \eqref{prob_penalty} with the objective function $g_{\sigma_k}(\bz)$. Suppose the positive sequence $\{\sigma_k\}$ is monotonically increasing and $\sigma_k\rightarrow +\infty$. Then any limit point of $\{\bz^k\}$ is a global minimizer of problem \eqref{prob}. 
\end{thm}

Theorem \ref{thm3} suggests that the penalty parameter $\sigma$ should go to infinity to guarantee that the obtained solution \{$x_{i,f}(k)$\} of \eqref{prob_penalty} is binary. In practice, however, the parameter $\sigma$ only needs to be large enough so that the values of the binary variables $\left\{x_{i,f}(k)\right\}$ are either close to zero or one. Then, a (feasible) binary solution of \eqref{prob} can be obtained by rounding.

Solving \eqref{prob_penalty} directly is not easy since it is a linearly constrained nonlinear program. To efficiently solve \eqref{prob_penalty}, we apply the SUM (Successive Upper bound Minimization) method \cite{Hunter2004,Razaviyayn2013}, which solves a sequence of approximate objective functions which are lower bounded by the original objective function. Due to the concavity of $P_\epsilon(\bx)$, the first order approximation of $P_\epsilon(\bx)$ is an upper bound of itself, i.e., $P_\epsilon(\bx)\leq P_\epsilon(\bx^t)+\nabla P_{\epsilon}(\bx^t)^T(\bx-\bx^t)$, where $\bx^t$ is the current iterate.
At the $(t+1)$-st iteration, we solve the following problem \vspace{-0.2em}
\begin{equation}\label{prob_sub}
\begin{array}{ll}
\hspace{-0.8em} \min\limits_{\br,\bx}
& g(\br)+\sigma_{t+1} \nabla P_{\epsilon}(\bx^{t})^T\bx\\
\hspace{-0.8em} \mbox{s.t.}&\eqref{link_capa}-\eqref{node_capa},\\
\hspace{-0.8em} & r_{ij}(k)\geq 0, r_{ij}(k,f)\geq 0,~\forall\,f,\forall\,k,~\forall\,(i,j)\in\cL,\\
\hspace{-0.8em} & x_{i,f}(k)\in [0,1],~\forall\,i\in V_f,\forall\,f\in \cF(k),\forall\,k. 
\end{array}
\end{equation}
Notice that each subproblem \eqref{prob_sub} is a linear program which can be efficiently solved to global optimality. 


{\bf Adding Additional Constraints.} The feasible region of the PSUM subproblem \eqref{prob_sub} is actually enlarged compared with that of the original problem \eqref{prob}, as the binary variables are relaxed while all the other constraints are unchanged.
To improve the feasibility of the obtained solution to the original problem, we will add some valid cuts. This idea is popular in combinatorial optimization.
To this end, we add some constraints related to the binary variables and strengthen the node capacity constraints, which are redundant for problem \eqref{prob} but can significantly reduce the feasible solution set of problem \eqref{prob_sub}.

We first define two new binary variables. Let $x_{i,f}$ be the binary variable indicating whether node $i$ provides function $f$ (i.e., $x_{i,f}=1$ if node $i$ provides $f$, otherwise $x_{i,f}=0$), and $x_i$ be the binary variable indicating whether node $i$ is active for providing network services (i.e., $x_{i}=1$ if node $i$ is active, otherwise $x_{i}=0$). By the definitions of binary variables, we have
\begin{equation}\label{xrelation}
  x_{i,f}(k) \leq x_{i,f} \leq x_i,~\forall\, i\in V_f,\,f\in\cF(k),\,k.
\end{equation}

Since the computational resource at each node $i$ is available only when node $i$ is active, the node capacity constraint \eqref{node_capa} can be strengthened in the following way.
\begin{equation}\label{node_capa1}
  \sum_f\sum_k\lambda(k)x_{i,f}(k)\leq x_{i}\mu_i,~\forall\,i.
\end{equation}
Moreover, for each function $f$ that node $i$ can provide, the computational resource at node $i$ is available for function $f$ only when the node provides function $f$. Therefore, the consumed computational resource on function $f$ at node $i$ is upper bounded by $x_{i,f}\mu_i$, as shown in equation \eqref{node_capa2}.
\begin{equation}\label{node_capa2}
  \sum_k\lambda(k)x_{i,f}(k)\leq x_{i,f}\mu_i,~\forall\,i\in V_f,\,f.
\end{equation}

With the above constraints added to problem \eqref{prob_sub}, the problem in the $(t+1)$-st PSUM iteration becomes
\begin{equation}\label{prob_sub2}
\begin{array}{ll}
\hspace{-0.8em} \min\limits_{\br,\bx}
& g(\br)+\sigma_{t+1} \nabla P_{\epsilon}(\bx^{t})^T\bx\\
\hspace{-0.8em} \mbox{s.t.}&\eqref{link_capa}-\eqref{node_capa}, \eqref{xrelation}-\eqref{node_capa2},\\
\hspace{-0.8em} & r_{ij}(k)\geq 0, r_{ij}(k,f)\geq 0,~\forall\,f,\forall\,k,~\forall\,(i,j)\in\cL,\\
\hspace{-0.8em} & x_i\in [0,1],~x_{i,f}\in [0,1],~\forall\, i,f,\\
\hspace{-0.8em} & x_{i,f}(k)\in [0,1],~\forall\,i\in V_f,\forall\,f\in \cF(k),\forall\,k. 
\end{array}
\end{equation}

The proposed PSUM algorithm for solving problem \eqref{prob} is presented in Algorithm \ref{table:PSUM}, where $\gamma$ and $\eta$ are two predefined constants satisfying $0<\eta<1<\gamma$.
\begin{algorithm}[h]
\caption{PSUM Algorithm for Solving Problem \eqref{prob}}
  0.~Solve problem \eqref{prob} with relaxed binary variables, and obtain solution $\bz^0=(\br^0,\bx^0)$;\\
  1.~Initialize $\epsilon_1, \sigma_1, T_{max}$, and let $t=0$;\\
  2.~{\bf While} $t< T_{max}$ {\bf Do}\\
  $\qquad$ Let $\sigma=\sigma_{t+1}$ and $\epsilon=\epsilon_{t+1}$;\\
  $\qquad$ Solve problem \eqref{prob_sub2} with the initial point being $\bz^t$, and obtain a solution $\bz^{t+1}=(\br^{t+1},\bx^{t+1})$;\\
  $\qquad$ If $\bx^{t+1}$ is binary, stop; \\
  $\qquad$ otherwise set $t=t+1$, and let $\sigma_{t+1}=\gamma \sigma_t, \epsilon_{t+1}=\eta\epsilon_t$;\\
  $\quad${\bf End}
  \label{table:PSUM}
\end{algorithm}
We remark that 1) the parameter $\epsilon$ is adaptively updated as the iteration number increases, which turns out to be very helpful in improving numerical performance of the algorithm \cite{Jiang2016}; 2) since the difference of two consecutive PSUM subproblems only lies in the objective function, the warm-start strategy can be applied in PSUM to solve the sequence of LP subproblems, i.e., let $\bz^t$ be the initial point of the $(t+1)$-st PSUM subproblem; 3) reference \cite{Shi2016} proposed a Penalty-BSUM algorithm that relaxes some equality constraints and applies Block-SUM to solve the penalized problem, while our Penalty-SUM algorithm is designed to enforce the relaxed variables being binary.

\subsection{PSUM-R: A Combination of PSUM and A Rounding Technique}\label{sec:rounding}

As we have mentioned before, in practice the penalty parameter $\sigma$ only needs to be sufficiently large so that the values of binary variables $\left\{x_{i,f}(k)\right\}$ are either close to zero or one.
To obtain a satisfactory solution and save computational efforts, we can obtain a suboptimal solution by the PSUM-R algorithm, using the PSUM algorithm with an effective rounding strategy. 

Suppose a solution of problem \eqref{prob} is obtained after a few PSUM iterations with some binary elements $\{\bigl(\bar{x}_{i,f}(k)\bigr)_{i\in V_f}\}$ being fractional.
We aim to construct a binary solution $\{x_{i,f}(k)\}$ on the basis of $\{\bigl(\bar{x}_{i,f}(k)\bigr)_{i\in V_f}\}$.
However, systematic rounding is difficult since the link and node capacity constraints \eqref{link_capa} and \eqref{node_capa}
couple the transmission of all $K$ flows. Thus we turn to round some elements to one while some elements to zero in a heuristic way.

Our strategy is to first obtain a binary solution of $\bx$, then solve the routing problem to obtain a solution of $\br$.
In particular, if $\bar\bx_{f}(k)$ is binary, then we simply let $\bx_f(k)=\bar\bx_f(k)$; for each $\bar\bx_f(k)$ that is not binary, our main idea of rounding is to respect the value of $\bx_f(k)$ (i.e., round the maximum element to one) when there exists one element being very close to one, otherwise to give priority to the node with the largest computational capacity.
In particular, we first check the value of its maximum element, if the maximum element is sufficiently close to one, then let the corresponding node provide function for flow $k$, that is, if $\bar{x}_{j,f}(k)=\max_{i\in V_f}\bar{x}_{i,f}(k)\geq \theta$ where $\theta\in(0,1)$ is a predefined positive threshold, then we set
\begin{equation}\label{rounding1}
x_{j,f}(k)=1,~\textrm{and } x_{i,f}(k)=0,~\forall\, i\in V_f\setminus\{j\};
\end{equation}
otherwise, we find the node $v\in V_f$ with the maximum remaining computational capacity, and let node $v$ provide function for flow $k$, i.e., set
\begin{equation}\label{rounding2}
x_{v,f}(k)=1,\textrm{ and } x_{i,f}(k)=0,~\forall\, i\in V_f\setminus\{v\}.
 \end{equation}
 In this way, we obtain a binary solution $\{x_{i,f}(k)\}$. Notice that the node capacity constraints may be violated after the rounding.

To determine the routing solution $\br$, we fix the values of binary variables $\{x_{i,f}(k)\}$ and
solve the original problem with objective function being $g+\tau\Delta$, and
modify the link capacity constraints by
\begin{equation}\label{new_link_capa}
  \sum_{k}r_{ij}(k)\leq C_{ij}+\Delta,~\forall\, (i,j)\in \cL,
  \end{equation}
where the new variable $\Delta$ is the maximum violation on link capacity and the positive constant $\tau$ is the weight of $\Delta$ in the objective function.
Finally, we obtain the solution of both the service instantiation and the routing. The PSUM-R algorithm is given in Algorithm \ref{table:PSUM-R}.

Moreover, after we obtain the solution, the violations of link and node capacity can be respectively measured by
\begin{equation}\label{vioLink}
\delta_{ij}=\max\{0,\sum\limits_kr_{ij}(k)-C_{ij}\},~\forall\, (i,j)\in \mathcal{L},
\end{equation}
\begin{equation}\label{vioNode}
\pi_i=\max\{0,\sum_k\sum_f\lambda(k)x_{i,f}(k)-\mu_i\},~\forall\,i. \vspace{-0.3em}
\end{equation}
If there is no violation on link and node capacity constraints, the obtained solution is feasible to problem \eqref{prob}. Practically, the obtained solution is satisfactory if the resource violation is sufficiently small.
We will show later in Section \ref{sec:numerical} that the PSUM-R algorithm can obtain a satisfactory solution with small resource violations within less time than PSUM.
\begin{algorithm}[h]
\caption{PSUM-R Algorithm for Solving Problem \eqref{prob}}
  1.~Perform $t_{max}$ PSUM iterations and obtain a solution $\bar{\mathbf{z}}=(\bar{\mathbf{r}},\bar{\mathbf{x}})$;\\
  2.~Generate a binary solution of $\bx$ on the basis of $\bar{\mathbf{x}}$ by the rounding steps \eqref{rounding1} and \eqref{rounding2};\\
  3.~Determine traffic routing with minimizing link capacity violation and obtain a solution of flow variable $\br$.
  \label{table:PSUM-R}
\end{algorithm}

\section{Low-Complexity Heuristic Algorithms}


In practice, service requests arrive randomly with little future information of network resources or other requests, and the network must be sliced as soon as a request arrives. An online algorithm with high competitive ratio is needed \cite{Karp1993}. In this section, we propose two low-complexity heuristic algorithms to solve problem \eqref{prob}, including one online algorithm (heuristic algorithm I).

The idea of our proposed low-complexity algorithms is to decouple the joint service instantiation and traffic routing problem \eqref{prob} into two separate problems. More specifically, the first step is to determine the set of instantiated nodes given a requested service/requested services, and the second step is to perform traffic routing given the selected instantiated nodes.
In particular, after fixing the values of the binary variables to determine the service instantiation, the remaining flow routing can be realized by solving problem \eqref{prob}, which is an LP problem. To ensure that the routing problem always has a feasible solution, we adopt a strategy similar to PSUM-R, i.e., modify the link capacity constraints by \eqref{new_link_capa} and set the objective function to $g+\tau\Delta$.

To determine which function node a service function should be instantiated for each request, Openstack scheduling \cite{openstack_schedule} provides a method of filtering and weighting. The filtering step selects eligible nodes in view of computational resource, CPU core utilization, and the number of running instances etc. The weighting step determines the instantiation node after computing a weighted sum cost for each eligible node.
Inspired by Openstack scheduling, we perform service instantiation in a similar way. In particular, for each service function in a SFC, we first select eligible function nodes in terms of their remaining computational resource, then compute the weighted sum cost of each eligible node $i$ by
\begin{equation}
  z_i=\sum_m w_m \times \textrm{ncost}(i,m),~\forall ~i,
\end{equation}
where ncost$(i,m)$ is the normalized cost of node $i$ in terms of capability $m$.
For example, if we consider the weighting factors of path length and available computational resource, the weighted sum cost can be computed by
\begin{equation}\label{eqn_weight}
  z_i=w_1(h^{S(k)}_i+h^{D(k)}_i)+w_2\frac{1}{\tilde\mu_i},
\end{equation}
where $h^{S(k)}_i$ (resp. $h^{D(k)}_i$) is the number of hops between nodes $i$ and $S(k)$ (resp. $D(k)$), $\tilde\mu_i$ is the current capacity of node $i$, and $w_1, w_2$ are weights.
After computing the weighted costs, we let the node with the smallest cost be the instantiation node.
Notice that there may be violations of node computational capacities.
In Algorithms \ref{table:heu_alg1} and \ref{table:heu_alg2}, we present heuristic algorithms I and II respectively.

{\bf Heuristic algorithm I.} We sequentially perform service instantiation and traffic routing for each flow (request).
As shown in Algorithm \ref{table:heu_alg1}, for each flow $k$, we first locate the nodes that provide functions in SFC $\cF(k)$ by filtering and weighting, next route flow $k$ from its source node to its destination node while going through the instantiation nodes in the specified order, then update node capacity and link capacity before proceeding to the next flow.

\begin{algorithm}[h]
\caption{Heuristic Algorithm I.}
  \textbf{For each flow $k$}\\
  $\quad$ \textbf{For each function $f^k_s$}\\
  $\quad$~ Select eligible function nodes with remaining computational capacity;\\
  $\quad$~ Determine the instantiation node by weighting using \eqref{eqn_weight};\\
  $\quad$~ Update remaining node capacity;\\
  $\quad$ \textbf{End}\\
  $\quad$ Route for flow $k$;\\
  $\quad$ Update link capacity: $C_{ij}=C_{ij}-r_{ij}(k),~\forall (i,j)$;\\
  \textbf{End}
  \label{table:heu_alg1} 
\end{algorithm}

{\bf Heuristic algorithm II.} We first perform service instantiation for all flows, then route all flows at the same time. Since the routing problem considers all flows, the violation on link capacity tends to be smaller than that in heuristic algorithm I.

\begin{algorithm}[h]
\caption{Heuristic Algorithm II.}
  \textbf{For each flow $k$}\\
  $\quad$ \textbf{For each function $f^k_s$}\\
  $\quad$~ Select eligible function nodes with remaining computational capacity;\\
  $\quad$~ Determine the instantiation node by weighting using \eqref{eqn_weight};\\
  $\quad$~ Update remaining node capacity;\\
  $\quad$ \textbf{End}\\
  \textbf{End}\\
  Route for all flows by solving an LP.
  \label{table:heu_alg2}
\end{algorithm}

\section{Numerical Experiments}\label{sec:numerical}

In this section, we present numerical results to illustrate the effectiveness of the proposed algorithms.
We shall compare our proposed PSUM algorithm, PSUM-R algorithm, and heuristic algorithms I and II with heuristic algorithm III (which is modified from the algorithm in \cite{Li2015}) and heuristic algorithm IV (which is proposed in \cite{Ghaznavi2016}).
We give the detailed description of heuristic algorithms III and IV in Appendix D.
All simulations are done in MATLAB (R2013b) with 2.30 GHz CPUs. All LP subproblems are solved by the optimization solver Gurobi 7.0.1 \cite{gurobi}.

To show the performance of these algorithms, we will compare the quality of the obtained solutions in terms of their objective function values, the CPU time, and the amount of violations of the network resource capacities. We define the worst-case violation ratio of link capacity by $\max_{ij}\frac{\delta_{ij}}{C_{ij}}$, and the worst-case violation ratio of node capacity by $\max_i\frac{\pi_i}{\mu_i}$, where $\delta_{ij}$ and $\pi_i$ are defined in \eqref{vioLink} and \eqref{vioNode} respectively.
We will consider two network topologies: a mesh topology in Fig. \ref{fig:topo1} and a fish topology in Fig. \ref{fig:topo2}.

In the PSUM algorithm, we set $T_{max}=20, ~\sigma_1=2, ~\epsilon_1=0.001, ~\gamma=1.1$, and $\eta=0.7$. In the PSUM-R algorithm, we set $t_{max}=7$ and $\theta=0.9$. In heuristic algorithms I and II, we set weights $w_1=1$ and $w_2=10\max_i\mu_i$ in \eqref{eqn_weight}. 

\subsection{Mesh Network Topology}\label{sec:topo1}

\begin{figure}[h]
  \centering
  \includegraphics[width=6cm]{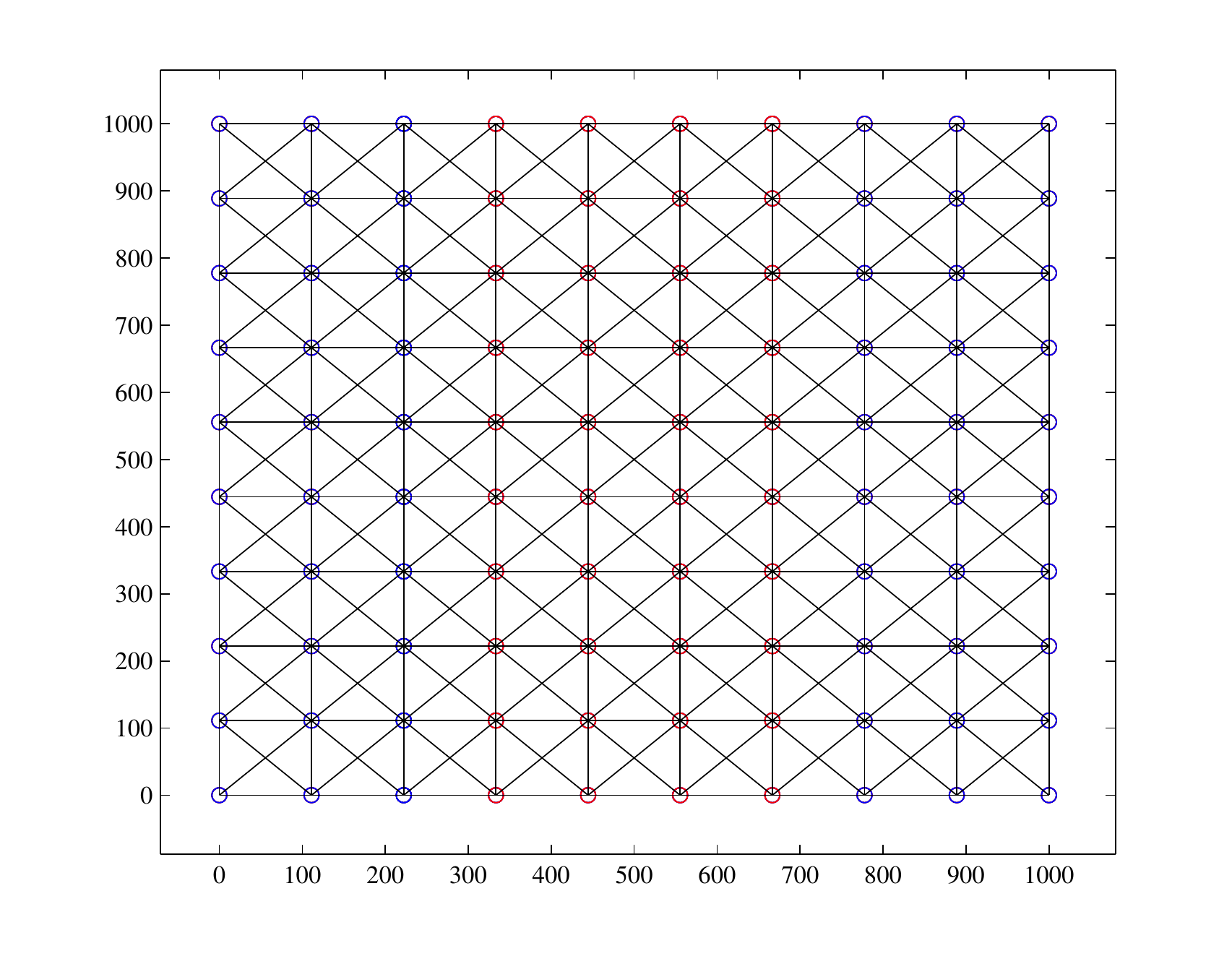}\vspace{-1em}
  \caption{\small Mesh topology.}
  \label{fig:topo1}
\end{figure}
We consider a mesh network with 100 nodes and 684 directed links. The topology of this network is shown in Fig. \ref{fig:topo1}, where the nodes located in the 4 middle columns are function nodes.
Suppose there are in total 5 service functions $\{f_1,\dots,f_5\}$ and 10 candidate function nodes for each function ($|V_f|=10$ for each function $f$).
We consider $K=30$ flows with demands $\lambda(k)=1$ for all flow $k$. The SFC $\cF(k)=(f^k_1\rightarrow f^k_2)$ and
$(S(k),D(k))$ are uniformly randomly chosen for each flow ($f^k_1\neq f^k_2, S(k),D(k)\notin V_{f^k_s},s=1,2$). The link capacity $C_{ij}$ is uniformly randomly chosen in $[0.5,5.5]$ and the node capacity $\mu_i$ is uniformly randomly chosen in $[0.5,8]$.
We randomly generate 50 instances of problem \eqref{prob} and apply the six algorithms to solve them.

\begin{figure}[h]
  \centering
  \includegraphics[width=7cm]{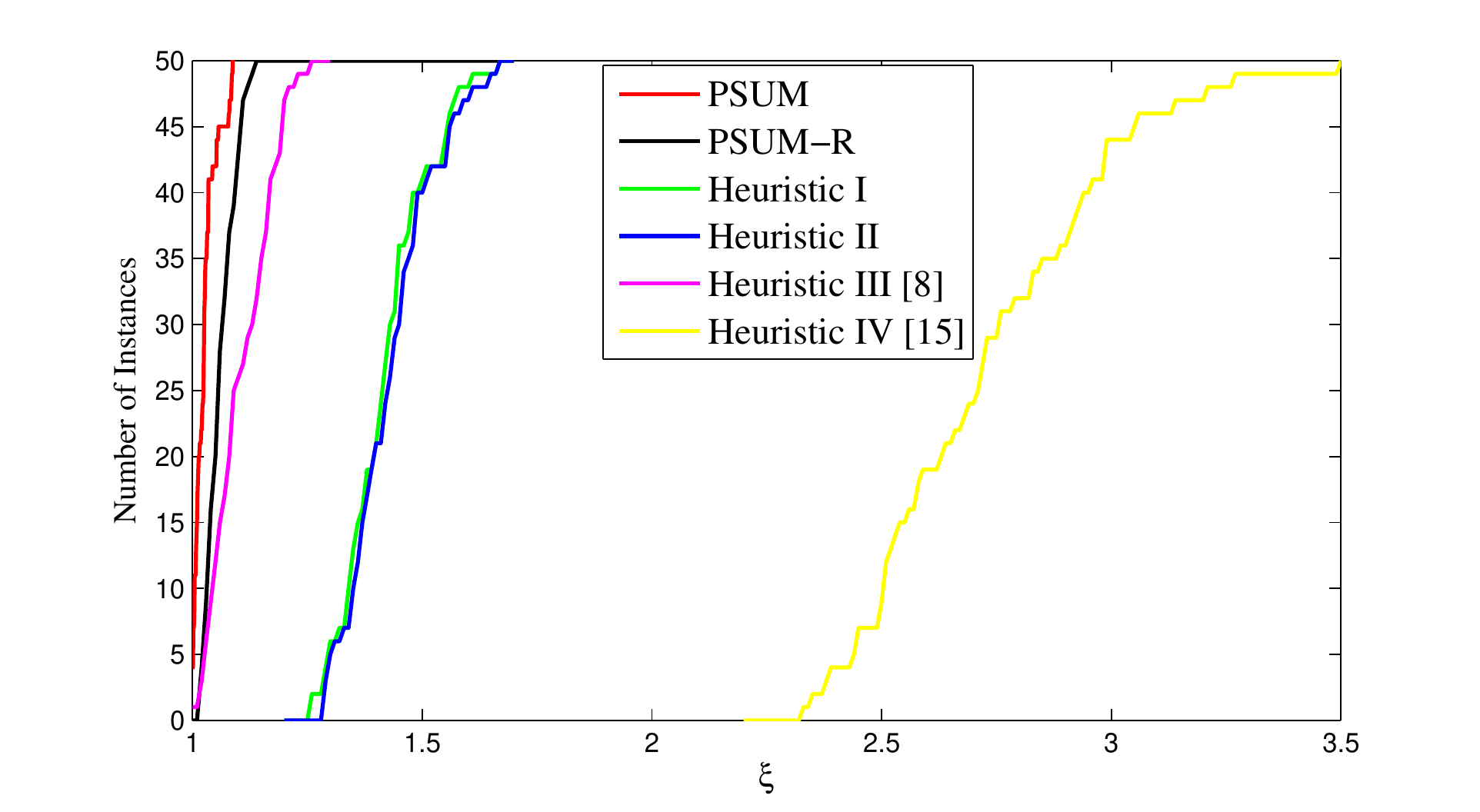}\vspace{-0.5em}
  \caption{\small [Mesh topology, Chain length=2] The number of instances where the ratio of the obtained objective value and the optimal value of the LP relaxation is less than or equal to $\xi\in[1,3.5]$.} \vspace{-1em}
  \label{fig:Val_topo1}
\end{figure}

Let $g_{\textrm{PSUM}}^*$ be the objective value of problem \eqref{prob} at the solution returned by the PSUM algorithm and let  $g_{\textrm{LP}}^*$ be the optimal value of the corresponding LP relaxation problem.
We find that $g^*_{\textrm{PSUM}}=g^*_{\textrm{LP}}$ in 4 instances and the objective value ratio $g^*_{\textrm{PSUM}}/g^*_{\textrm{LP}}$ is less than 1.09 in all of 50 instances, which implies that the solutions returned by the PSUM algorithm are close to be global optimal. We also find that in 21 instances the PSUM-R algorithm already obtains a feasible solution before the rounding step, which further shows that the PSUM algorithm converges fast.
As shown in Fig. \ref{fig:Val_topo1}, the objective value ratios corresponding to the PSUM-R algorithm and heuristic algorithm III are in [1.0, 1.26], the ratios corresponding to heuristic algorithms I and II are in [1.25, 1.7), and those corresponding to heuristic algorithm IV are in [2.2, 3.5] which are much larger. In terms of the returned objective value, the PSUM algorithm gives the best solutions and heuristic algorithm IV performs the worst among the six algorithms.
\begin{figure}[h]
  \centering
  \hspace{-1em}\includegraphics[width=4.7cm]{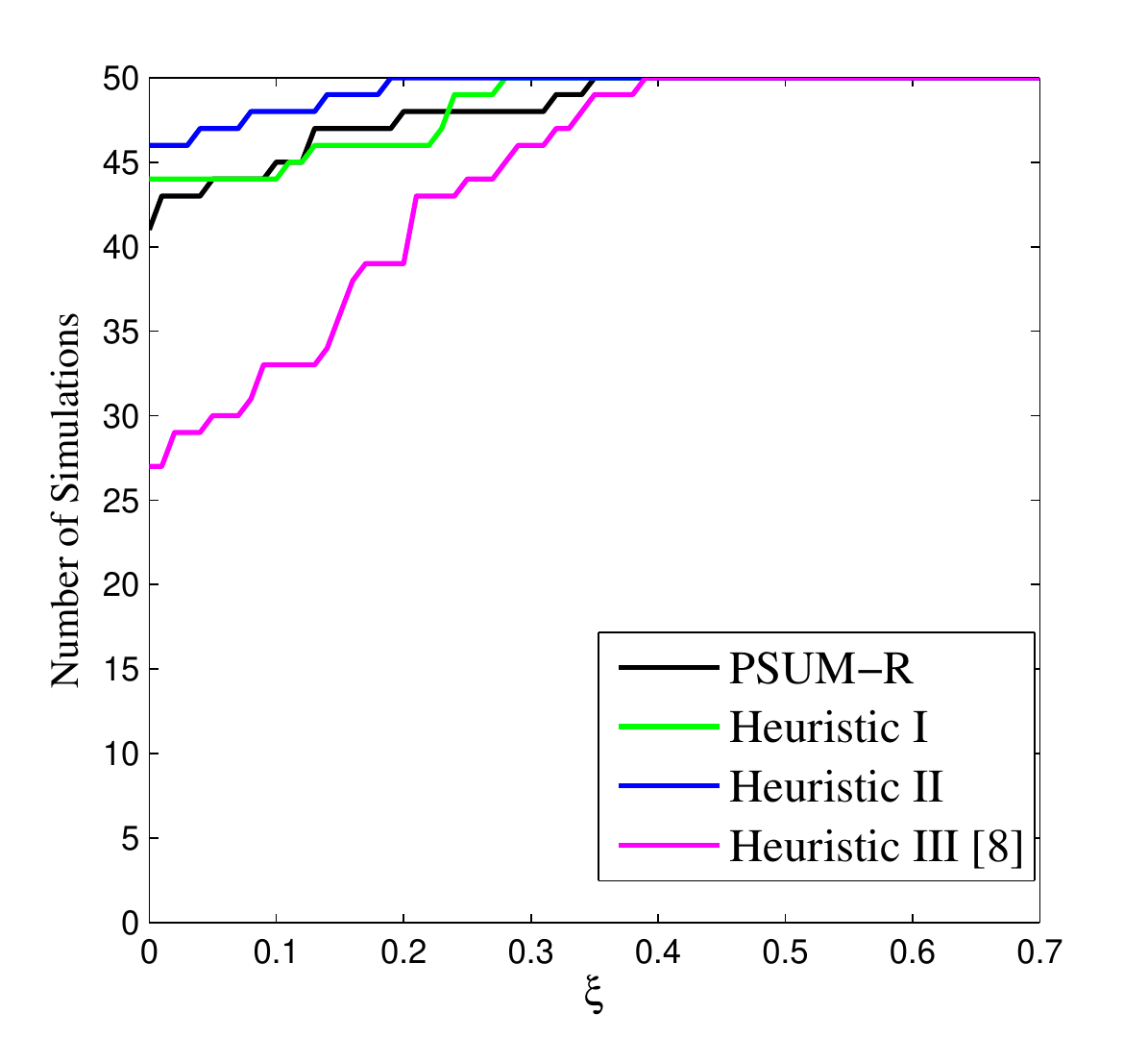}\hspace{-1em}
  \includegraphics[width=4.7cm]{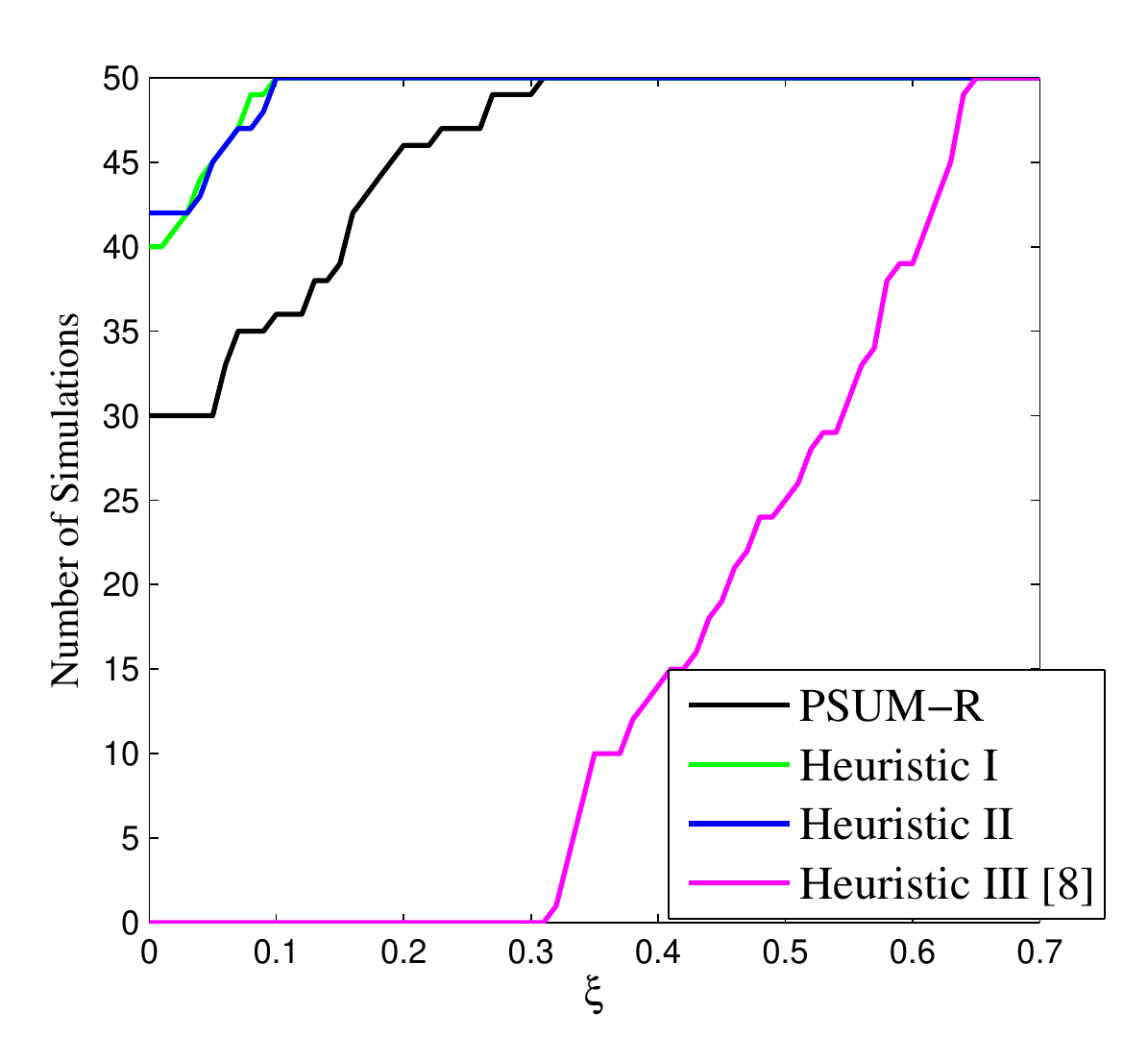} \vspace{-1em}
\caption{\small [Mesh topology, Chain length=2] The statistics of constraint violation ratios returned by PSUM-R and heuristic algorithms I, II, and III. Left: violation ratios of link capacity; Right: violation ratios of node capacity.} 
\label{fig:vio_topo1}
\end{figure}

The statistics of worst-case violation ratios are plotted in Fig. \ref{fig:vio_topo1}. Since there is no resource violation in the solutions obtained by the PSUM algorithm and heuristic algorithm IV, we only show the violation ratios returned by PSUM-R, heuristic algorithms I, II, and III.
From Fig. \ref{fig:vio_topo1} we can see that the violation ratios returned by heuristic algorithms I and II are small, and those returned by heuristic algorithm III are generally much larger.
Moreover, the violation ratios of node capacity returned by heuristic algorithms I and II are smaller than those returned by PSUM-R and heuristic algorithm III, as the node capacity has already been taken into consideration in the service instantiation step in our proposed heuristic algorithms I and II.

\subsection{Fish Network Topology}\label{sec:topo2}
\begin{figure}[h]
  \centering
  \includegraphics[width=6cm]{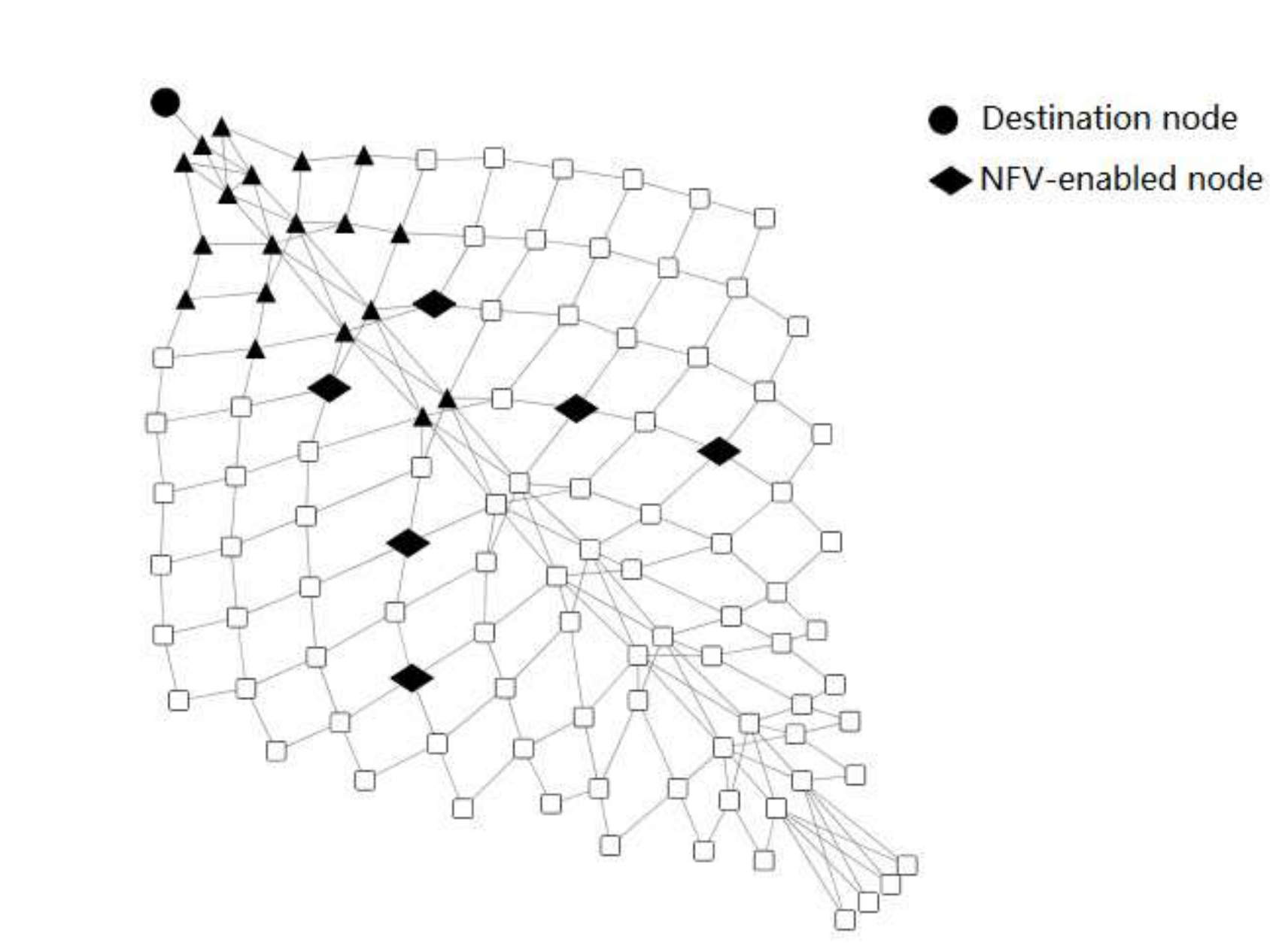}\vspace{-0.5em}
  \caption{\small Fish topology.}\label{fig:topo2}
\end{figure}
We consider another network with 112 nodes and 440 directed links. The topology of this network is shown in Fig. \ref{fig:topo2}, where the 6 diamond nodes are function nodes and the triangular nodes cannot be chosen as a source node of any flow.
We consider $K=20$ flows with a common destination which is the circular node in Fig. \ref{fig:topo2}.
According to the distance (counted by the number of hops) to the destination node, the network nodes are divided into 11 layers.
We set the node capacity $\mu_i$ to 16 for any function node $i$, and choose the value of link capacity $C_{ij}$ from [1, 55] which is
further divided into 10 sub-intervals.
In particular, the capacities of the links connecting layer $m+1$ and layer $m$ are uniformly randomly chosen from the $(11-m)$-th
sub-interval. Suppose there are in total 4 service functions $\{f_1,\dots,f_4\}$ and all of them can be provided by any of the 6 function
nodes.                                                                                                                                   
In each instance of problem \eqref{prob}, the demand $\lambda(k)$ is an integer number randomly chosen in $[1,5]$, the source node $S(k)$ is randomly chosen from the available network nodes excluding the function nodes, and the SFC $\cF(k)$ is an ordered sequence of functions randomly chosen from $\{f_1,\dots,f_4\}$ for each flow.

\begin{figure}[h]
  \centering
  \hspace{-1em}\includegraphics[width=4.7cm]{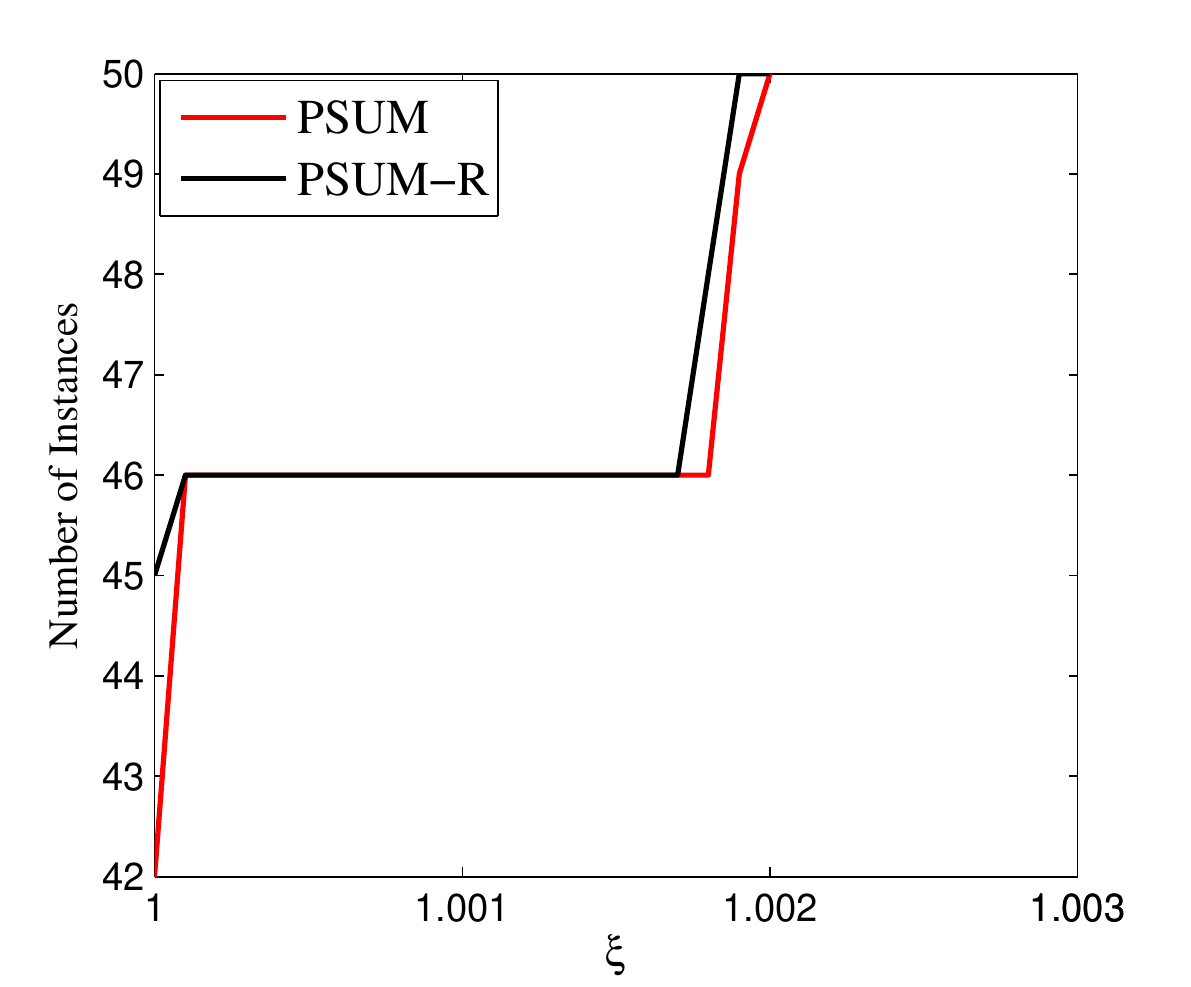}\hspace{-1em}
  \includegraphics[width=4.7cm]{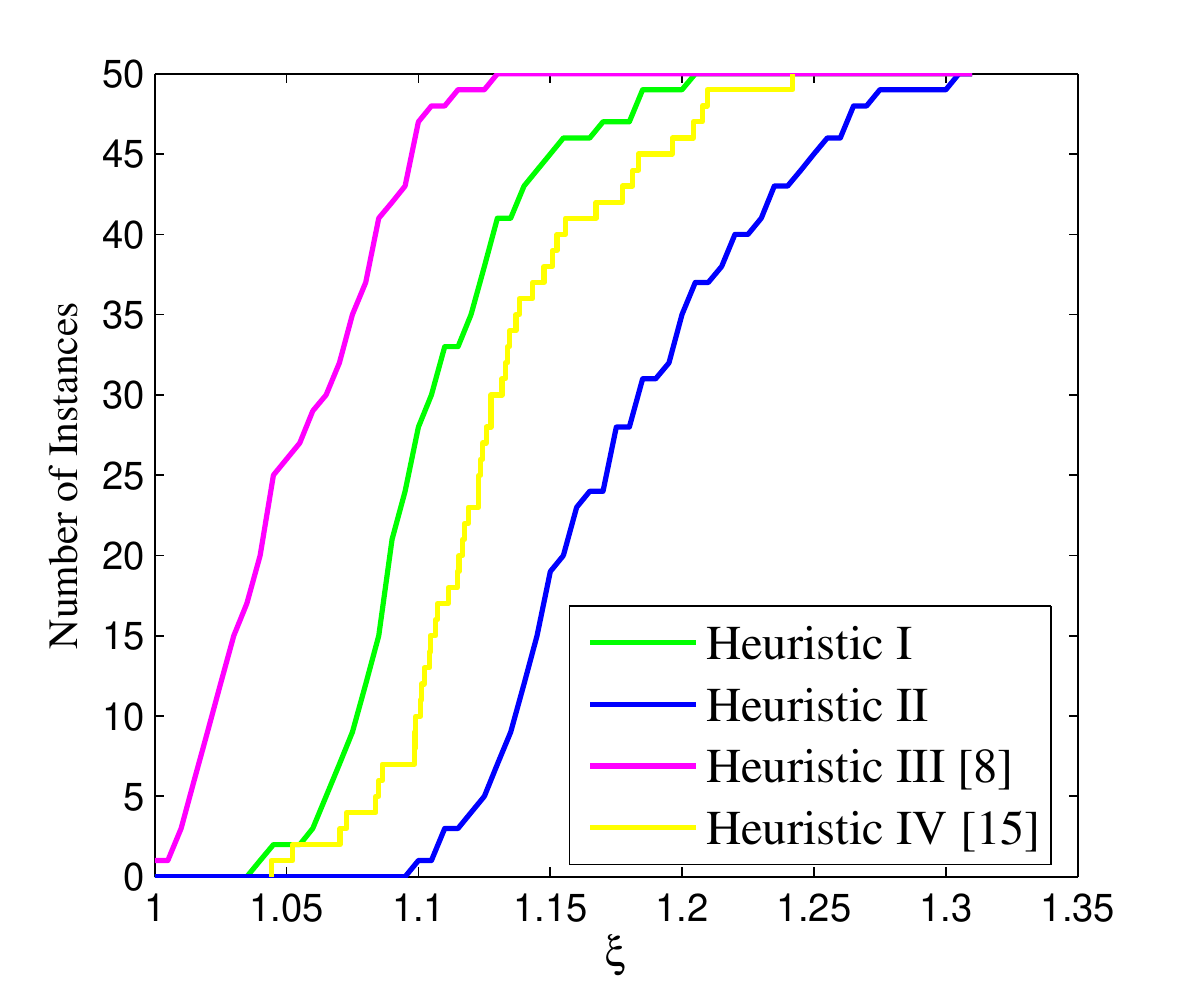} \vspace{-0.8em}
  \caption{\small [Fish topology, Chain length=1] The number of instances where the ratio of the obtained objective value and the optimal value of the LP relaxation is less than or equal to $\xi\in[1,1.31]$.}  \vspace{-1em}
   \label{fig:Val_case1}
\end{figure}

We first set the chain length to 1, i.e., each SFC $\cF(k)$ consists of only one service function, and randomly generate 50 instances of problem \eqref{prob}.
We find that  $g^*_{\textrm{PSUM}}=g^*_{\textrm{LP}}$ in 42 instances and $g^*_{\textrm{PSUM-R}}=g^*_{\textrm{LP}}$ in 45 instances. This implies that LP relaxation is tight for at least 42 instances (in the sense that the LP relaxation of problem \eqref{prob} and problem \eqref{prob} have same optimal values) and the PSUM algorithm finds global optimal solutions for these instances. Moreover, the objective value ratio $g^*_{\textrm{PSUM}}/g^*_{\textrm{LP}}$ is less than 1.002 in all of 50 instances, which implies that the obtained solution is very close to the global optimal solution.
As shown in Fig. \ref{fig:Val_case1}, heuristic algorithm III generates better solutions in terms of the objective value among the heuristic algorithms.
\begin{figure}[h]
  \centering
  \hspace{-1em}\includegraphics[width=4.7cm]{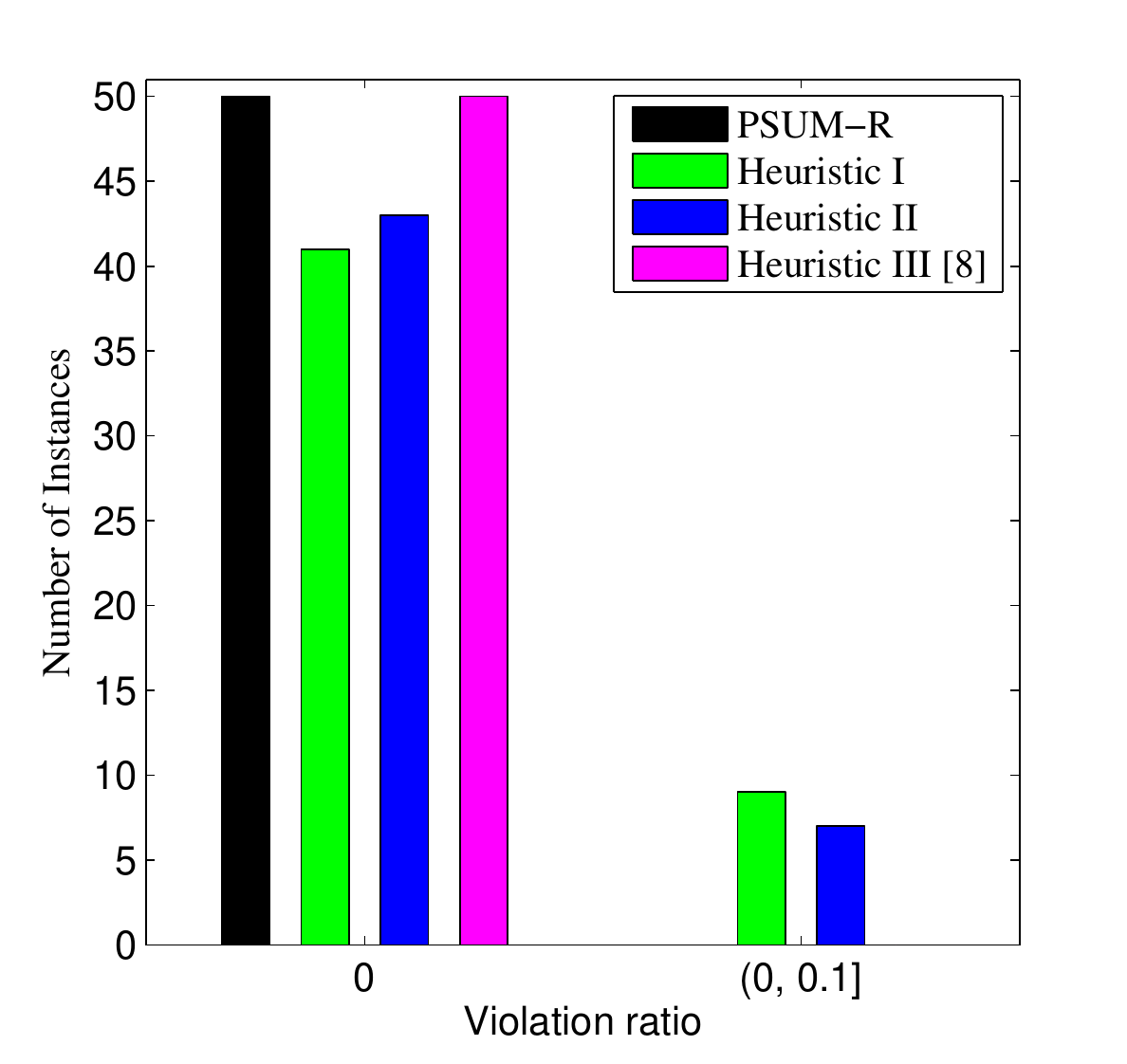}\hspace{-1em}
  \includegraphics[width=4.7cm]{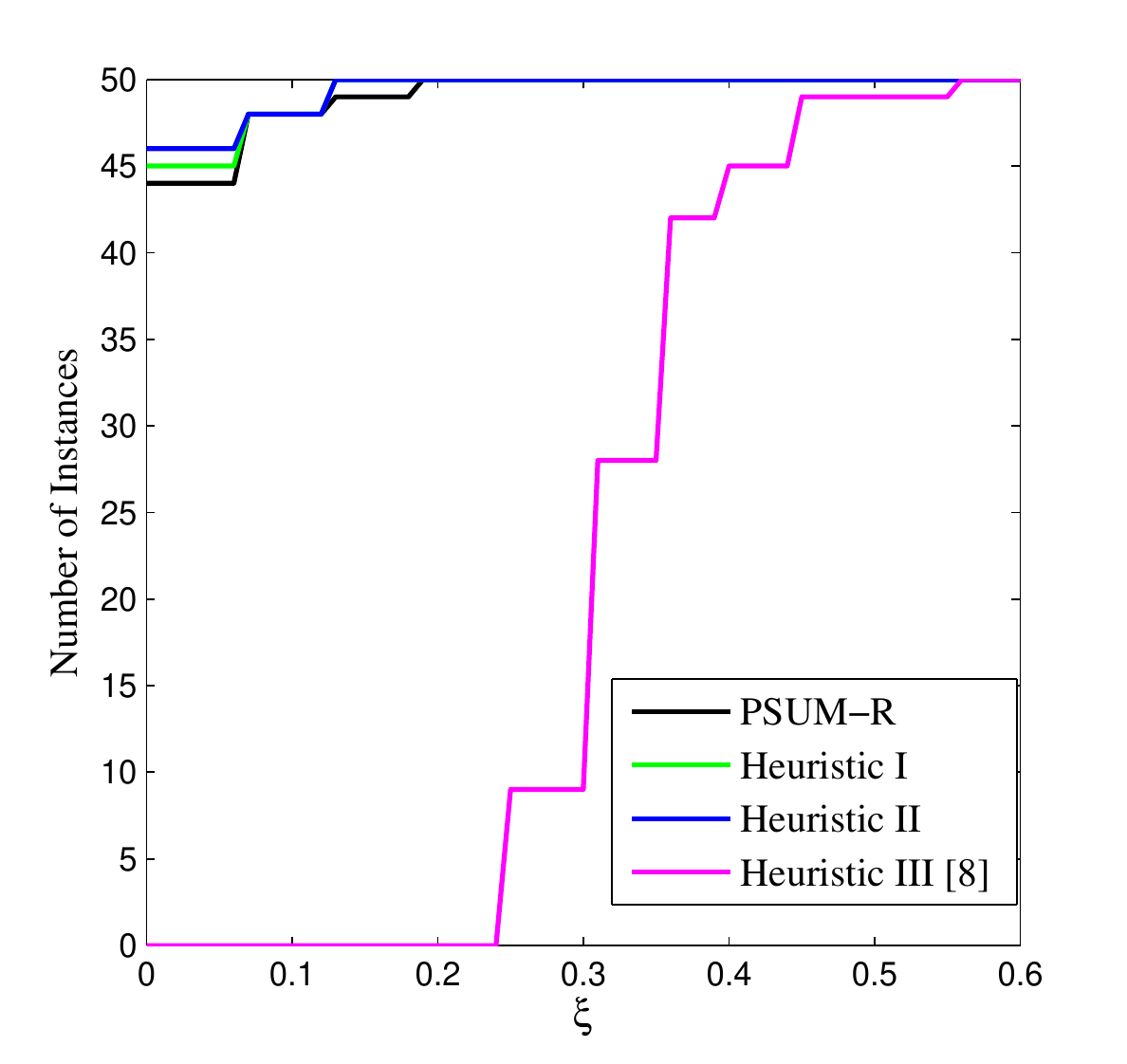} \vspace{-0.5em}
\caption{\small [Fish topology, Chain length=1] The statistics of constraint violation ratios. Left: violation ratios of link capacity; Right: violation ratios of node capacity.} 
\label{fig:vio_case1}
\end{figure}
The statistics of worst-case violation ratios are plotted in Fig. \ref{fig:vio_case1}. 
We can see from Fig. \ref{fig:vio_case1} that there is no link/node capacity violation in the solutions obtained by PSUM-R and heuristic algorithms I and II in over 40 simulations, while larger violations on node capacity in the solutions obtained by heuristic algorithm III.

\begin{figure}[h]
  \centering
  \includegraphics[width=6cm]{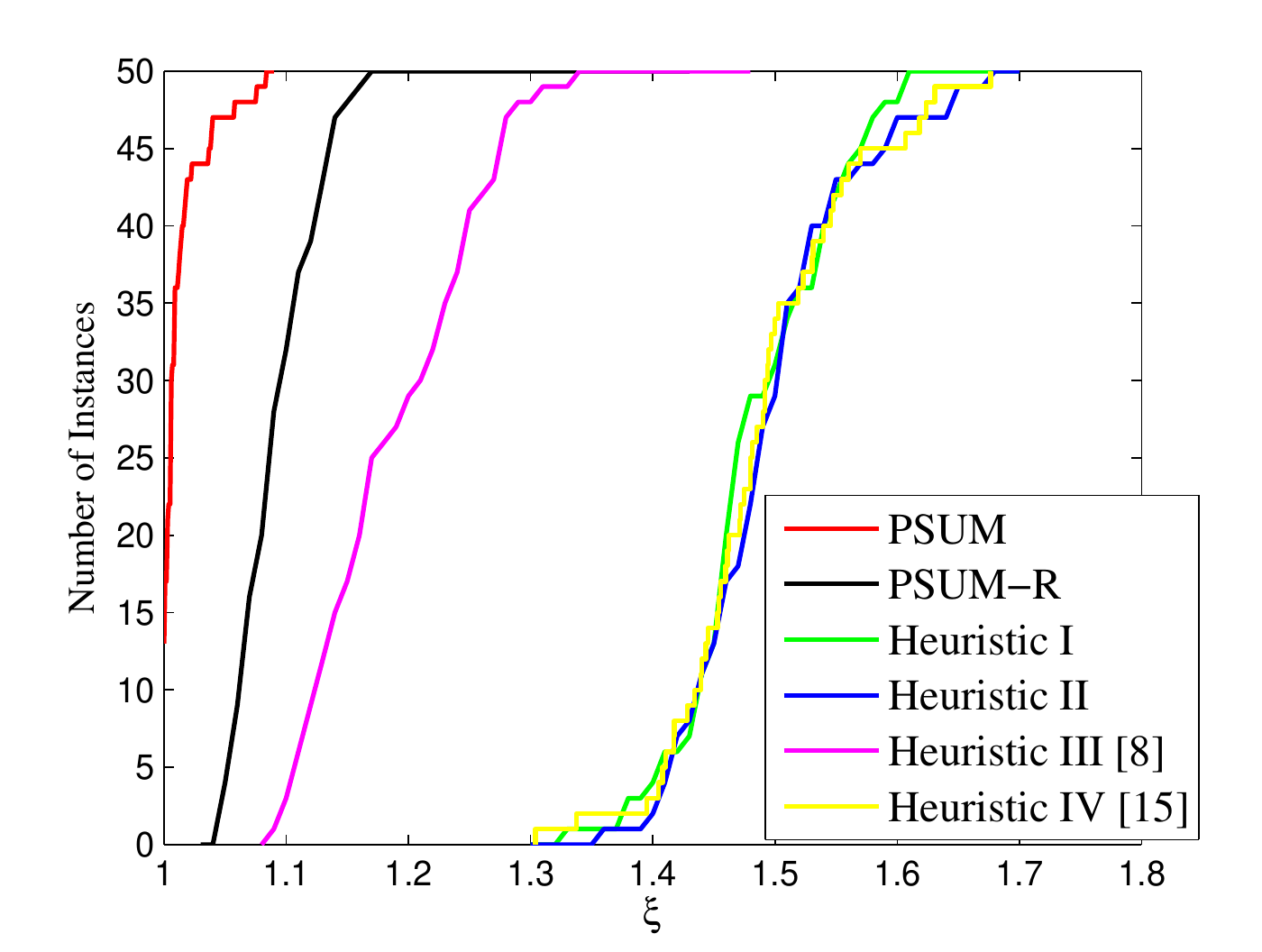}\vspace{-1em}
  \caption{\small [Fish topology, Chain length=2] The number of instances where the ratio of the obtained objective value and the optimal value of the LP relaxation is less than or equal to $\xi\in[1,1.7]$.} \vspace{-1em}
  \label{fig:Val_case2}
\end{figure}

\begin{figure}[h]
  \centering
  \hspace{-1em}\includegraphics[width=4.6cm]{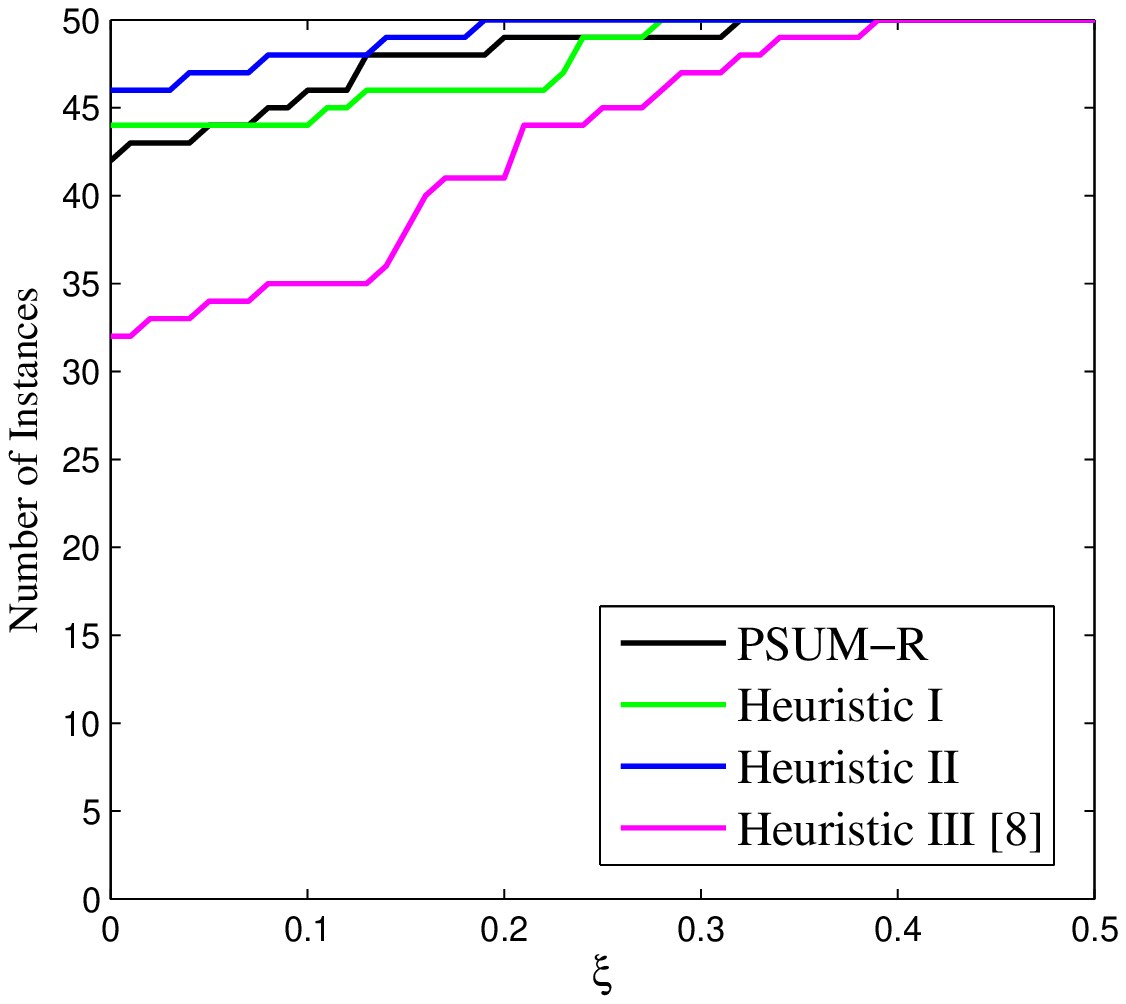}\hspace{-1em}
  \includegraphics[width=4.6cm]{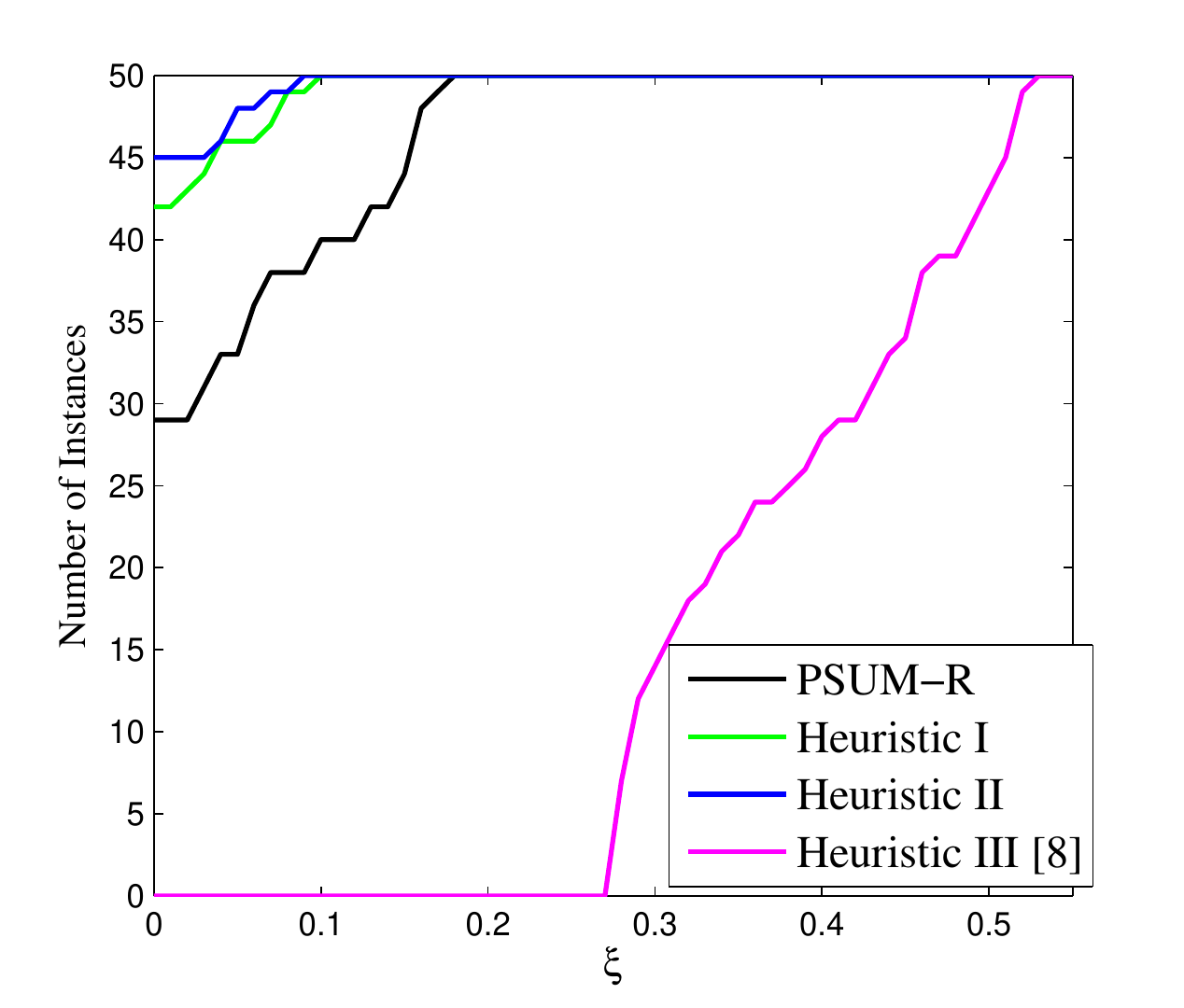} \vspace{-1em}
\caption{\small [Fish topology, Chain length=2] The statistics of constraint violation ratios. Left: violation ratios of link capacity; Right: violation ratios of node capacity.}\vspace{-1em}
\label{fig:vio_case2}
\end{figure}

We set the chain length to 2, i.e., each SFC $\cF(k)$ consists of two different service functions.
Fig. \ref{fig:Val_case2} shows the statistics of the objective value ratios.
In this case, the ratio $g^*_{\textrm{PSUM}}/g^*_{\textrm{LP}}$ is below 1.09 in all 50 instances and is below 1.01 in 30 instances, and the ratios returned by heuristic algorithms IV are in [1.3, 1.7], which are very similar to those returned by heuristic algorithms I and II. One reason for the better performance of heuristic algorithm IV in this topology (compared to the mesh topology in Section \ref{sec:topo1} where the chain length is also 2) is that, the multi-layer routing problem solved in heuristic IV becomes less complicated when $|V_f|$ decreases. Fig. \ref{fig:vio_case2} plots the violation ratios, from which we can see that the violation ratios become larger (compared to the case where the chain length is 1), especially for heuristic algorithm III.

From the results above we find that the performance of PSUM, PSUM-R and all four heuristic algorithms deteriorates when the chain length increases from 1 to 2. As shown in Theorem \ref{thm2}, if the link and node capacity are fixed, the LP relaxation of problem \eqref{prob} and itself are less likely to be equivalent as the chain length increases, and problem \eqref{prob} becomes more difficult to solve. Therefore, the gap between $g^*_{\textrm{LP}}$ and $g^*_{\textrm{PSUM}}$ tends to become larger when the chain length increases.

\begin{figure}[h]
  \centering
  \hspace{-1em}\includegraphics[width=4.4cm]{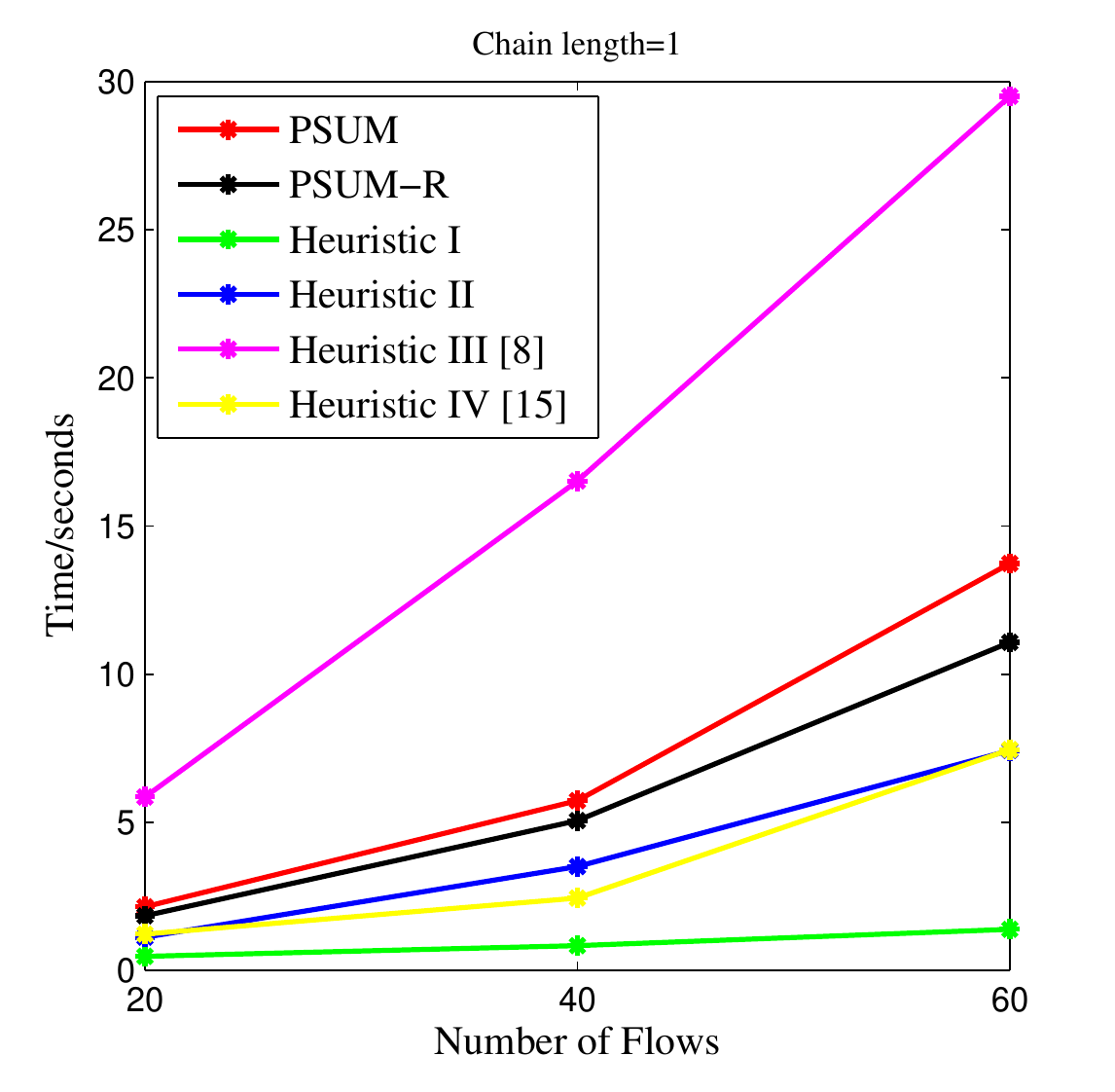}\hspace{-1em}
  \includegraphics[width=4.4cm]{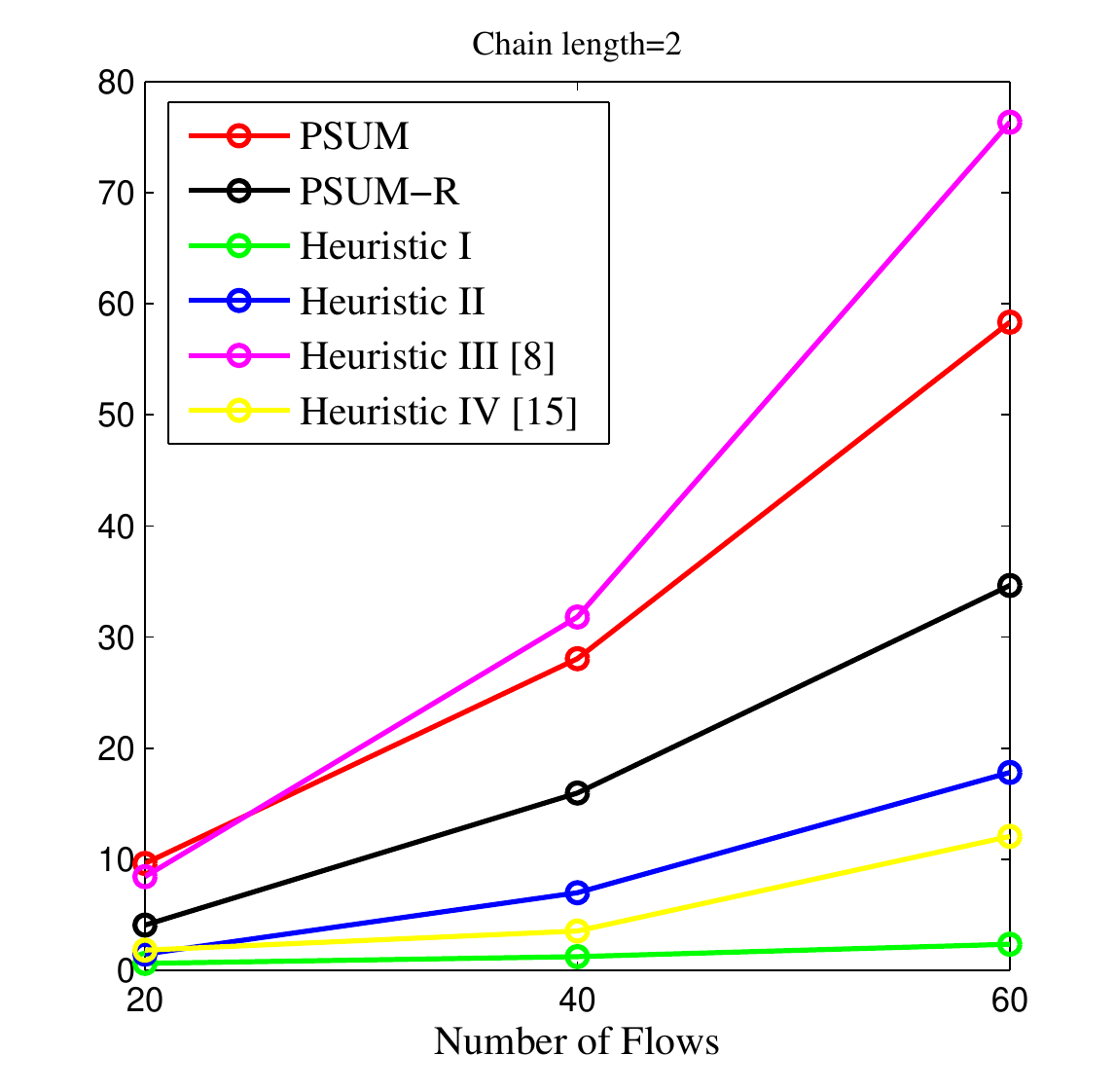} \vspace{-0.5em}
\caption{\small The CPU time (per instance) versus problem size. Left: [Fish topology, Chain length=1]; Right: [Fish topology, Chain length=2].}\vspace{-1em}
\label{fig:cputime}
\end{figure}

Finally, we plot the CPU time (in seconds) per instance with different problem sizes in Fig. \ref{fig:cputime}.
Among the six algorithms, heuristic algorithms I and IV are the fastest ones as the LP problems to be solved are in small scales.
The PSUM algorithm requires more time. Meanwhile, the time of PSUM-R (which performs 7 PSUM iterations) is about 80\% of the time of PSUM (whose maximum number of iterations is set to 20), which means the number of iterations performed in PSUM to obtain a feasible solution is usually much smaller than 20.
Heuristic algorithm III is more time-consuming than PSUM. This is because heuristic algorithm III usually needs to solve more LP problems and the warm-start strategy cannot be used in solving the LP problems.

\subsection{Summary of Simulation Findings}
From the above simulation results we can conclude that:
\begin{itemize}
\item Our proposed heuristic algorithms I and II are efficient and can find satisfactory solutions with moderate resource violations;
\item Heuristic III gives better solutions than heuristic algorithms I, II, and IV in terms of objective values, but has larger resource violations and is more time-consuming;
\item Heuristic algorithm IV can find feasible solutions with satisfactory objective values, but its performance is not stable under different numerical settings;
\item Our proposed PSUM algorithm guarantees perfect satisfaction of resource constraints and returns the best solution among all algorithms, but is slower than the heuristic algorithms I, II, and IV (albeit it is faster than heuristic algorithm III);
\item Our proposed PSUM-R algorithm achieves a good balance of solution quality and algorithm efficiency.
\end{itemize}

\section{Conclusion}\label{sec:conclusion}

In this work, we study the network slicing problem.
Different from most of the existing works, we assume that each flow receives any service function in the corresponding service function chain at exactly one function node, a requirement that is strongly motivated by reducing practical coordination overhead. We formulate the problem as a mixed binary linear program and prove its strong NP-hardness.
To effectively solve the problem, we propose a PSUM algorithm, a variant PSUM-R algorithm, and two low-complexity heuristic algorithms, all of which are easily implemented.
Our simulation results demonstrate that PSUM and PSUM-R can approximately solve the problem by returning a solution that is close to the optimal solution. Moreover, the PSUM algorithm completely respects the resource capacity constraints, and the PSUM-R algorithm achieves a good balance of solution quality and algorithm efficiency. 

\section*{Acknowledgment}

The authors would like to thank Navid Reyhanian for his help in numerical simulations.

\bibliographystyle{IEEEtran}
\bibliography{IEEEabrv,ref_SDRA}

\newpage
\appendices
\def\thesection{\Alph{section}}
\section{Proof of Theorem \ref{thm1}}
We will prove that for an instance of problem \eqref{prob}, the problem of checking its feasibility is as hard as a 3-dimensional matching problem, which is known as strongly NP-complete.

\begin{figure}[h]
  \centering
  \includegraphics[width=6cm]{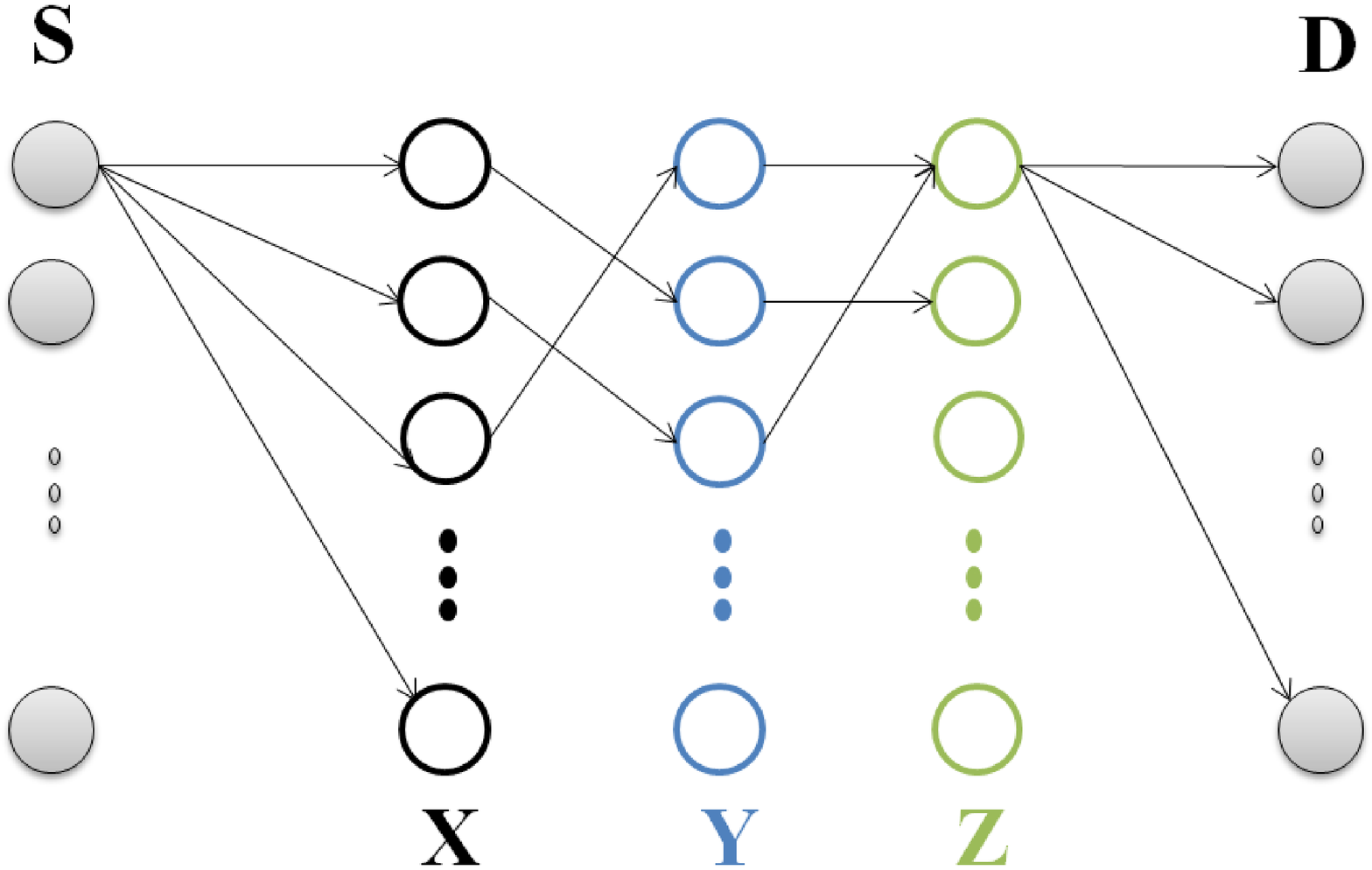}\vspace{-0.5em}
  \caption{Constructed network $\cN=(\cV,\cL)$ to prove Theorem 1}\vspace{-0.5em}
  \label{3match}
\end{figure}

We first construct an instance of problem \eqref{prob} as follows.
\begin{itemize}
\item The set of service functions: $F=F_1\cup F_2\cup F_3$, where $F_1,F_2,F_3$ are disjoint.
\item The SFC of flow $k$: $\cF(k)=(f^k_1\rightarrow f^k_2\rightarrow f^k_3)$, where $f^k_j\in F_j,~j=1,2,3,\,k=1,\dots,K$. Moreover, the SFCs of any two flows are different in the sense that $f^k_j\neq f^m_j$ for all $ k\neq m,\,j=1,2,3$.
\item The set of network nodes: $\cV=\cS\cup X\cup Y\cup Z\cup\cD$, where $\cS$ is the set of source nodes, $\cD$ is the set of destination nodes, $X,Y,Z$ are disjoint sets of function nodes, and these sets have the same number of elements, i.e., $K=|\cS|=|X|=|Y|=|Z|=|\cD|$. See Fig. \ref{3match} for an illustration. Moreover,
    any node $x\in X$ can provide all functions in $F_1$, and for any function $f\in F_1$, $V_f=X$;
    any node $y\in Y$ can provide all functions in $F_2$, and for any function $f\in F_2$, $V_f=Y$;
    any node $z\in Z$ can provide all functions in $F_3$, and for any function $f\in F_3$, $V_f=Z$.
\item The set of links consists of 3 parts: $\cL=\big\{(s,x)\mid s\in\cS, \,x\in X\big\}\cup \cL_R\cup\big\{(z,d)\mid z\in Z,\,d\in\cD\big\}$, where $\cL_R$ is determined by a set $R\subseteq X\times Y\times Z$ in the sense that any $(x,y,z)\in R$ implies that there exist directed links $(x,y)\in \cL_R, (y,z)\in \cL_R$.
\item All flows have the same rate, i.e., $\lambda(k)=\lambda$ for all $k$.
\item The link capacity $C_{ij}$ is sufficiently large that $C_{ij}\geq 4K\lambda$ for any $(i,j)\in\cL$.
\item For any function node $i\in X\cup Y\cup Z$, the value of $\mu_{i}$ is chosen to ensure that node $i$ provides at most one function, e.g., $\mu_i\in[\lambda, 2\lambda)$, which ensures that function node $i$ provides at most one function due to the limited node capacity.
\end{itemize}

In the following, we prove that the constructed above instance of problem \eqref{prob} has a feasible solution
if and only if there exists a 3-dimensional matching of $R$.
\begin{itemize}
\item[(i)] problem \eqref{prob} has a feasible solution $\Longrightarrow$
there exists a 3-dimensional matching of $R$.

Since problem \eqref{prob} has a feasible solution, then for each flow $k$,
there exist function nodes $x^k\in X, y^k\in Y, z^k\in Z$ such that $x^k$
provides $f^k_1$, $y^k$ provides $f^k_2$, $z^k$ provides $f^k_3$, and
$(x^k,y^k),(y^k,z^k)$ are directed links in $\cL_R$.

Let us define a subset of $X\times Y\times Z$ by $M=\{(x^k,y^k,z^k)\mid
k=1,\dots,K\}$. By the definition of $\cL_R$, we have $M \subseteq R$.
Moreover, for any $(x^k,y^k,z^k), (x^t,y^t,z^t)\in M~(k\neq t)$, we have
$x^k\neq x^t, y^k\neq y^t,z^k\neq z^t$. This is because that each node
provides at most one function. Therefore, $M$ is a 3-dimensional matching of
$R$.

\item[(ii)] $M\subseteq R$ is a 3-dimensional matching $\Longrightarrow$
problem \eqref{prob} has a feasible solution.

Since $M$ is a 3-dimensional matching of $R$, $M$ consists of $K$ triplets.
We can build a one-to-one mapping between the $K$ flows and the $K$ triplets in
$M$. We denote the triplet that flow $k$ is mapped to as $(x^k,y^k,z^k)$.
Let $x^k$ provide $f^k_1$, $y^k$ provide $f^k_2$, and $z^k$ provide $f^k_3$ for
any flow $k$. Since $(x^k,y^k,z^k)\in R$, $(x^k,y^k),(y^k,z^k) \in \cL$ and
for any $t\neq k$ we have $x^k\neq x^t, y^k\neq y^t,z^k\neq z^t$. Thus this
instantiation strategy is feasible. The routing problem is also feasible
because the link capacity is sufficiently large. In this way, we obtain a
feasible solution of problem \eqref{prob}. 
\end{itemize}

We can see obviously from the above description that the construction of
this instance can be completed in polynomial time. Since the 3-dimensional
matching problem is strongly NP-complete, finding a feasible solution of \eqref{prob}
is also strongly NP-complete, and thus solving problem \eqref{prob} itself is strongly
NP-hard. The proof is completed.

\section{Proof of Theorem \ref{thm2}}
By the constraint $\sum\limits_{f\in \cF(k)} x_{i,f}(k)\leq 1$ and the node capacity constraint $\sum_f\sum_k\lambda(k) x_{i,f}(k)\leq \mu_{i}$, we have
\[\sum_k\sum_f\lambda(k)x_{i,f}(k)\leq \sum_k\lambda(k)\cdot 1=\bar\mu.\]
By the relationship between $r_{ij}(k)$ and $r_{ij}(k,f)$ and the fact that
$r_{ij}(k,f)\leq \lambda(k)$ for all $(i,j),f\in\cF(k)$, we have
\[\sum_kr_{ij}(k)=\sum_k\sum_{f\in \cF(k)\cup{\{f^k_0\}}}r_{ij}(k,f)\leq \sum_k\lambda(k)(|\cF(k)|+1)=\bar C.\]
Therefore, the link and node capacity constraints are automatically
satisfied if $\mu_i\geq \bar\mu$ for all $i$ and $C_{ij}\geq \bar C$ for all
$(i,j)$. Then the LP relaxation of problem \eqref{prob} reduces to
\begin{equation}\label{prob2}
\begin{array}{ll}
\mbox{minimize}& \displaystyle\sum\limits_{(i,j)}\sum_{k}\sum_{f\in \cF(k)\cup{\{f^k_0\}}} r_{ij}(k,f)\\\vspace{+0.5em}
\mbox{subject to}&{\sum\limits_{i\in V_f} x_{i,f}(k)=1}, ~\forall\,k,f\in \cF(k),\\\vspace{+0.5em}
& \sum\limits_{f\in F(k)} x_{i,f}(k)\leq 1, ~\forall\,k,i,\\\vspace{+0.5em}
& \textrm{flow conservation constraints},\\\vspace{+0.5em}
& r_{ij}(k,f)\geq0,~x_{i,f}(k)\in [0,1],~\forall \,k,f,i,\\ 
\end{array}
\end{equation}
which decouples among different flows. Thus, solving problem \eqref{prob2}
is equivalent to solving $K$ subproblems, and the $k$th subproblem aims at
minimizing the total link rates in directing flow $k$ from its source node
to its destination node while sequentially going through the instantiation
nodes in the order of the functions in the chain $\cF(k)$. Let us denote
$\mathbb{P}_k$ as the set of the shortest paths from $S(k)$ to $D(k)$ that
sequentially go through nodes that provide $f^k_1,f^k_2,\dots,f^k_n$. We
have the following claim which can be proved by contradiction.

{\bf Claim}: The optimal solution of problem \eqref{prob2} for each flow $k$
is to transmit with rate of $\lambda(k)$ (or $\lambda(k)$ amount of flow)
along any path in $\mathbb{P}_k$.

In fact, we can always find an optimal solution of problem \eqref{prob2} that has
binary components of $\{x_{i,f}(k)\}$. For each flow $k$, we select a path
from $\mathbb{P}_k$ and an instantiation node on the path for each required
service function (this is feasible due to the definition of $\mathbb{P}_k$),
and let all data of flow $k$ be transmitted on this path. Then we obtain a
solution of problem \eqref{prob2}. According to the above claim, such
solution must be an optimal solution of problem \eqref{prob2}. Therefore,
problem \eqref{prob} has an optimal solution with binary components of
$\{x_{i,f}(k)\}$, which implies the first conclusion in Theorem 2.

Next, we prove by contradiction that the bounds in \eqref{bounds} on $\left\{\mu_i, C_{ij}\right\}$ are tight, i.e., there exists an instance of problem \eqref{prob} where some $\mu_i$ or $C_{ij}$ is below the bound and the LP relaxation problem does not have an optimal solution with components of $\{x_{i,f}(k)\}$ being binary.
\begin{figure}[h]
  \centering
  \includegraphics[width=8.5cm]{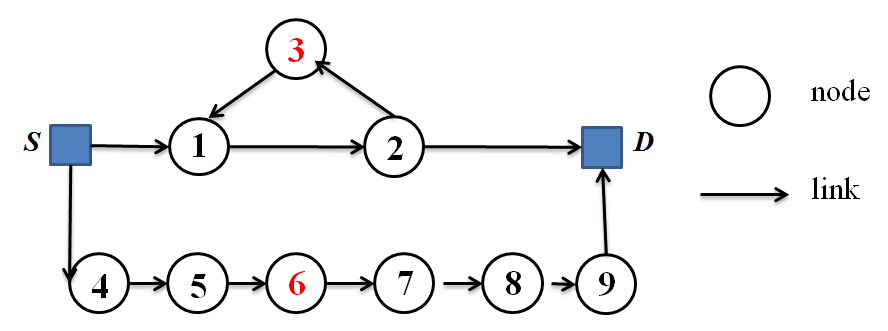}\\
  \caption{Considered network to prove the tightness of the bounds in \eqref{bounds}}
  \label{thm2proof0}
\end{figure}

Let us consider the network shown in Fig. \ref{thm2proof0}. Suppose there is one flow with demand $\lambda=1$, the source node is $S$ and the destination node is $D$, and the service function chain is $\cF=\left\{f_1\right\}$.
In this network, only nodes 3 and 6 can provide $f_1$. By \eqref{bounds}, we can compute the bounds for this case: $\bar\mu=\lambda=1$,  $\bar C=2\lambda=2$.

\begin{figure}[h]
  \centering
  \subfigure[]{
  \includegraphics[width=6cm]{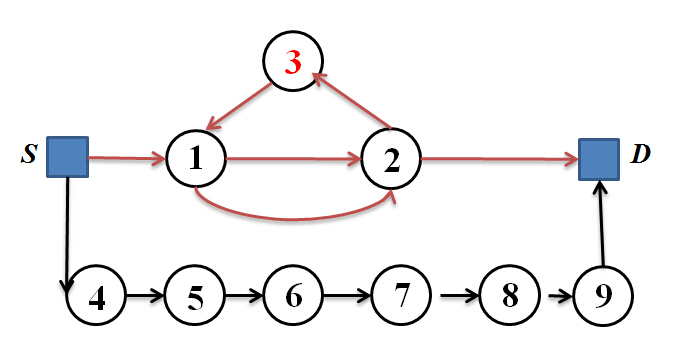}}
  \subfigure[]{
  \includegraphics[width=6cm]{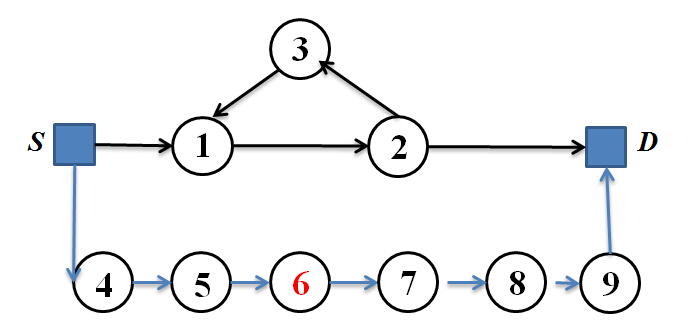}}\vspace{-0.5em}
  \caption{}\vspace{-0.5em}
  \label{pf_instance}
\end{figure}

Since $V_{f_1}=\{v_3, v_6\}$, we have that the shortest path $\mathbb{P}_1=\{(S, v_1, v_2, v_3, v_1, v_2, D)\}$ with the number of hops being 6 (see the path in red in Fig. \ref{pf_instance} (a)), and another feasible path is $(S, v_4, v_5, v_6, v_7, v_8, v_9, D)$ with the number of hops being 7 (see the path in blue in Fig. \ref{pf_instance} (b)). To minimize the total-link-rate objective function in problem \eqref{prob2}, the flow should transmit as many data as possible on the shortest path $\mathbb{P}_1$.

(i) Let $\mu_3=1-\epsilon<\bar \mu$ where $\epsilon\in(0,1)$,
$\mu_6>\bar\mu$, and $C_{ij}$ is no less than $\bar C$ for any link $(i,j)$.
Since $v_3$ is on the shortest path and $\mu_3=1-\epsilon$, $v_3$ can
process at most $1-\epsilon$ units of data. Therefore, in the optimal
solution of the LP relaxation problem, $1-\epsilon$ units of data are transmitted
on the shortest path $\mathbb{P}_1$, while the remaining $\epsilon$ units of
data are transmitted on the other feasible path. The optimal solution of
$\{x_{i,f}(1)\}$ is \emph{not} binary.

(ii) Let $C_{12}=2(1-\epsilon)<\bar C$, the capacity of other links are
larger than $\bar C$, and the capacity of any function node is larger than
$\bar\mu$.
Since link $(1,2)$ is on the shortest path, by the same analysis in (i) we
have that the amount of data transmitted on the shortest path is at most
$1-\epsilon$. The remaining $\epsilon$ units of data must be transmitted on
the other feasible path. Therefore, the optimal solution of $\{x_{i,f}(1)\}$
of the LP relaxation problem is \emph{not} binary.

From the above two cases we can conclude that the lower bounds of
$\{\mu_i,C_{ij}\}$ given in \eqref{bounds} are tight. The proof is completed.

\section{Proof of Theorem \ref{thm3}}\label{sec:penalProof}
For ease of presentation, we define $P^k=P_{\epsilon}(\bx^k)$ and $g^k=g(\br^k)$.
Since $\bz^k$ is a global minimizer of problem \eqref{prob_penalty} with the objective function $g_{\sigma_k}(\bz)$, it follows that 
\begin{equation*}
  g_{\sigma_k}(\bz^k)\leq g_{\sigma_k}(\bz^{k+1}),~ g_{\sigma_{k+1}}(\bz^{k+1})\leq g_{\sigma_{k+1}}(\bz^k),~\forall\,k. 
\end{equation*}
Combining the above with the assumption $\sigma_{k}\leq \sigma_{k+1}$, we obtain 
\begin{equation*}
\sigma_k(P^k-P^{k+1})\leq g^{k+1}-g^k\leq  \sigma_{k+1}(P^k-P^{k+1}),~\forall\,k, 
\end{equation*}
which shows that \{$g^k$\} is increasing and \{$P^k$\} is decreasing.

Suppose that $\bz^*$ is a global minimizer of problem \eqref{prob}. Then $P_\epsilon(\bz^*)=0$. By the definition of $\bz^k$, we have
$g_{\sigma_k}(\bz^k)\leq g_{\sigma_k}(\bz^*)=g(\bz^*)$, which further implies that \vspace{-0.2em}
\begin{equation}\label{eq1}
0\leq g^k+\sigma_k P^k\leq g(\bz^*). 
\end{equation}
This, together with the facts that $g^k\geq 0, P^k\geq 0$, and $\sigma_k\rightarrow +\infty$, shows that $\sigma_kP^k\rightarrow 0$ and $P^k\rightarrow 0$ as $k\rightarrow +\infty$.

Let $\bar \bz=(\bar\br,\bar\bx)$ be any limit point of $\{\bz^k\}$, and $\{\bz^k\}_\cK$ be a subsequence converging to $\bar\bz$. Since $P^k\rightarrow 0$, we have $P_{\epsilon}(\bar\bx)=0$, which shows that $\bar\bz$ is feasible for \eqref{prob}. Furthermore, taking limit along $\cK$ in \eqref{eq1}, we have $g(\bar\bz)\leq g(\bz^*)$. Therefore, $g(\bar\bz)=g(\bz^*)$ and $\bar \bz$ is a global minimizer of \eqref{prob}.$\Box$

\section{Description of Heuristic Algorithms III and IV }
As shown in Algorithm \ref{alg:heu3} below, heuristic algorithm III is modified from the heuristic algorithm in \cite{Li2015}.
In this modified algorithm, we denote the set of binary variables $\{x_{i,f}(k)\}$ as $\cB$, the set of $\{x_{i,f}(k)\}$ which take value of one as $\cB_1$, and those taking value of zero as $\cB_0$. The basic idea is to first determine the value of the binary variable by ``bootstrapping iteration" and ``greedy selection" and then perform traffic routing.
\begin{algorithm}[h]
\textbf{Bootstrapping iteration:}\\
   $\quad${\bf For} $t=1:t_{max}$\\
  $\qquad$ Solve problem \eqref{prob} with relaxed binary variables and with $x_{i,f}(k)\in \cB_1$ being fixed to be one. \\
  $\qquad$ Let the solution be \{$x^*_{i,f}(k)$\} and \\
  $\qquad$ ~~$\cB'_1=\{(i,f,k)\mid x^*_{i,f}(k)\geq \theta_2\}$, $\cB'_0=\{(i,f,k)\mid x^*_{i,f}(k)\leq \theta_1\}$;\\
  $\qquad$ Update $\cB'_1$ by checking node capacity constraints, i.e., let $x_{i,f}(k)=1$ for all $(i,f,k)\in \cB'_1$,\\
  $\qquad$  ~~and check whether $\sum_f\sum_{k:(i,f,k)\in \cB'_1}\lambda(k)x_{i,f}(k)\leq \mu_i$ holds for all $i$, remove those \\
  $\qquad$  ~~from $\cB'_1$ that occur in the violated inequalities;\\
  $\qquad$ Let $\cB_1=\cB'_1,~\cB_0=\cB'_0$, and $\cB'=\cB\setminus(\cB_1\cup \cB_0)$.\\
  $\quad${\bf End}\\
  \textbf{Greedy selection:}\\
  $\quad$ For each $(i,f,k)\in \cB'$, solve the LP relaxation problem with $x_{i,f}(k)=0$. If the problem is infeasible, add the index into $\cB_1$; \\
  $\quad$ For $x_{i,f}(k)$ whose being assigned to zero leads to the maximum decrease or least increase in the objective, add the index into $\cB_0$; \\
  \textbf{Rounding technique:}\\
  $\quad$ For the variables in $\cB'$, determine the value by the rounding technique in Section III-B in \cite{sdra2017};\\
  $\quad$ Solve the problem with binary variables being fixed, and measure the link and node capacity violations.
  \caption{Heuristic Algorithm III.}
  \label{alg:heu3}
\end{algorithm}

Heuristic algorithm IV is proposed in \cite{Ghaznavi2016}, for which we describe in Algorithm \ref{alg:heu4}.
This heuristic algorithm reduces solving problem \eqref{prob} to solving a sequence of subproblems which are defined between two consecutive layers.
A layer $f$ is defined as a set of function nodes in $V_f$ (i.e., the set of nodes that can provide function $f$), and layer $f^k_0$ (resp. $f^k_{n+1}$) refers to the source (resp. destination) node of flow $k$.
For each flow $k$, we will route it between layers to bring the traffic from the first layer ($S(k)$) to the last layer $(D(k))$.
In particular, for each subproblem defined between layer $f^k_s$ and layer $f^k_{s+1}$, we first determine the instantiation of function $f^k_{s+1}$ for flow $k$ by solving a multidimensional knapsack problem,
next solve a multi-source multi-sink Minimum Cost Flow (MCF) problem to route the traffic,
and finally perform local search to improve the obtained solution (for example, change the instantiation node to see whether the cost can be reduced).

\begin{algorithm}[h]
  $\quad${\bf For} flow $k=1:K$\\
  $\qquad${\bf For} $s=0:n-1$\\
  $\qquad\quad$ Determine the instantiation node of function $f^k_{s+1}$ for flow $k$ by heuristically \\
  $\qquad\qquad$ solving a knapsack problem;\\
  $\qquad\quad$ Route flow $k$ from layer $f^k_s$ to layer $f^k_{s+1}$ by solving a MCF problem;\\
  $\qquad\quad$ Improve the solution by local search;\\
  $\qquad$ {\bf End}\\
  $\qquad\quad$ Route flow $k$ from layer $f^k_n$ to the destination node $D(k)$ by solving a MCF problem;\\
  $\quad${\bf End}
  \caption{Heuristic Algorithm IV.}
  \label{alg:heu4}
\end{algorithm}

%



\end{document}